\documentclass[12pt]{article}

\usepackage{graphicx}
\usepackage{color}

\newcommand\res{PNASlowres} 

\newcommand{\ket}[1]{|#1 \rangle}  
\newcommand{\bra}[1]{\langle #1 |}  
\newcommand{\ketbra}[2]{\left|#1\rangle\!\langle #2\right|}
\newcommand{\text}[1]{\mathrm{#1}}
\newcommand{\braket}[2]{\langle #1 | #2 \rangle}
\newcommand{\average}[1]{\langle #1 \rangle}
\newcommand{\Prj}[1]{|#1 \rangle\!\langle #1|}
\newcommand{\mc}[1]{\mathcal{#1}} 

\newcommand{\one}{{1}}
\newcommand{\ii}{{\rm i}}

\newcommand{\ee}{{\rm e}}
\newcommand{\eg}{\textit{e.g.}}
\newcommand{\ie}{\textit{i.e.}}
\newcommand{\etc}{\textit{etc.}}

\newcommand{\La}{$^{1\!}L_\mathrm{a}$}
\newcommand{\Lb}{$^{1\!}L_\mathrm{b}$}

\newcommand{\eqref}[1]{\textup{eq.~{\ref{#1}}}}
\newcommand{\appropto}{\mathrel{\vcenter{
  \offinterlineskip\halign{\hfil$##$\cr
    \propto\cr\noalign{\kern2pt}\sim\cr\noalign{\kern-2pt}}}}}
    
\newcommand{\Heff}{\hat{H}_{\mathrm{eff}}}
\newcommand{\ER}{\mc{E}^\textrm{R}}
\newcommand{\EL}{\mc{E}^\textrm{L}}
\newcommand{\CR}{C^\textrm{R}}
\newcommand{\CL}{C^\textrm{L}}
\newcommand{\U}{\hat{U}}
\newcommand{\UU}{\hat{\mathcal{U}}}
\newcommand{\tr}{\mbox{tr}}

\usepackage{scicite}

\usepackage{times}

\newenvironment{sciabstract}{
\begin{quote} \bf}
{\end{quote}}

\newcounter{lastnote}
\newenvironment{scilastnote}{
\setcounter{lastnote}{\value{enumiv}}
\addtocounter{lastnote}{+1}
\begin{list}
{\arabic{lastnote}.}
{\setlength{\leftmargin}{.22in}}
{\setlength{\labelsep}{.5em}}}
{\end{list}}

\title{Ultraviolet superradiance from mega-networks of tryptophan in biological architectures}

\author{N.~S.~Babcock,$^{1}$ G.~Montes-Cabrera,$^{1,2}$ K.~E.~Oberhofer,$^{3}$ \\ M.~Chergui,$^{3}$ G.~L.~Celardo,$^{4}$ P.~Kurian$^{1\ast}$ \\
\\
\normalsize{$^{1}$Quantum Biology Laboratory, Howard University, Washington DC, USA} \\
\normalsize{$^{2}$Institute of Physics, Benemérita Universidad Autónoma de Puebla, Mexico} \\
\normalsize{$^{3}$Lausanne Centre for Ultrafast Science, École Polytechnique Fédérale de Lausanne, Switzerland} \\
\normalsize{$^{4}$Department of Physics and Astronomy, CSDC and INFN, Florence Section, University of Florence, Italy} \\
\\
\normalsize{$^\ast$To whom correspondence should be addressed; E-mail:  pkurian@howard.edu.}
}

\date{}

\begin{document}

\baselineskip24pt

\maketitle

\begin{sciabstract}
Networks of tryptophan --- an aromatic amino acid with strong fluorescent response --- are ubiquitous in biological systems, forming diverse architectures in transmembrane proteins, cytoskeletal filaments, sub-neuronal elements, photoreceptor complexes, virion capsids, and other cellular structures. We analyze the cooperative effects induced by ultraviolet (UV) excitation of several biologically relevant tryptophan mega-networks, thus giving insight into novel mechanisms for cellular signalling and control. Our theoretical analysis in the single-excitation manifold predicts the formation of strongly superradiant states 
due to collective interactions among organized arrangements of up to more than $10^5$ tryptophan UV-excited transition dipoles in microtubule architectures, 
which leads to an enhancement of the fluorescence quantum yield that is confirmed by our experiments. We demonstrate the observed consequences of this superradiant behavior in the fluorescence quantum yield for hierarchically organized tubulin structures, which increases in different geometric regimes at thermal equilibrium before saturation --- highlighting the effect's persistence in the presence of disorder. 
\end{sciabstract}

\section*{Introduction}

Tryptophan is the only amino acid with an indole moiety, 
making it a suitable precursor for a number of metabolites involved in biological signalling, most notably kynurenine and the neurotransmitter serotonin \cite{cervenka_kynurenines_2017}, which share tryptophan's highly aromatic character. 
It is an ideal fluorescent reporter of biomolecular dynamics, given its natural occurrence in proteins, its strong ultraviolet absorption, and its significant absorption-emission Stokes shift that is highly sensitive to the protein, solvent, and electrostatic environments. 
As a matter of fact, in recent years tryptophan has been used as a reporter of the Stark effect in photoactivated proteins \cite{Schenkl05, leonard_functional_2009}, to monitor protein folding kinetics \cite{tusell_simulations_2012}, as the operative chromophore in  
resonance energy transfer networks of UV-specific photoreceptor complexes \cite{li_leap_2020, li_dynamics_2022}, as a reporter of charge-transfer states in proteins \cite{callis_quantitative_2004, consani_ultrafast_2013} and of solvation dynamics at lipid-water and protein-water interfaces \cite{lu_femtosecond_2004, xu_picosecond_2015}, to track local electrostatic changes in diverse classes of proteins \cite{vivian_mechanisms_2001}, and as a probe for conformational ensembles of proteins in solution \cite{pan_correlation_2011}, among other applications.

Tryptophan residues are often found in transmembrane proteins, situated at the lipid-water interface. Multi-tryptophan proteins have been widely studied, including myoglobin, hemoglobin, cytochrome-\textit{c} oxidase, and cytochrome \textit{P}-450 \cite{khan_steady-state_1997}, as well as in the photoreceptors cryptochrome \cite{lin2018circadian}, bacteriorhodopsin \cite{Schenkl05, leonard_functional_2009}, and UVR8 \cite{li_leap_2020, li_dynamics_2022}. 
Large, organized tryptophan networks occur in these transmembrane proteins, receptors, and other macromolecular aggregates, lending essential structural and functional value to living systems.

\begin{figure}[ht]
\includegraphics[width=.9\linewidth]{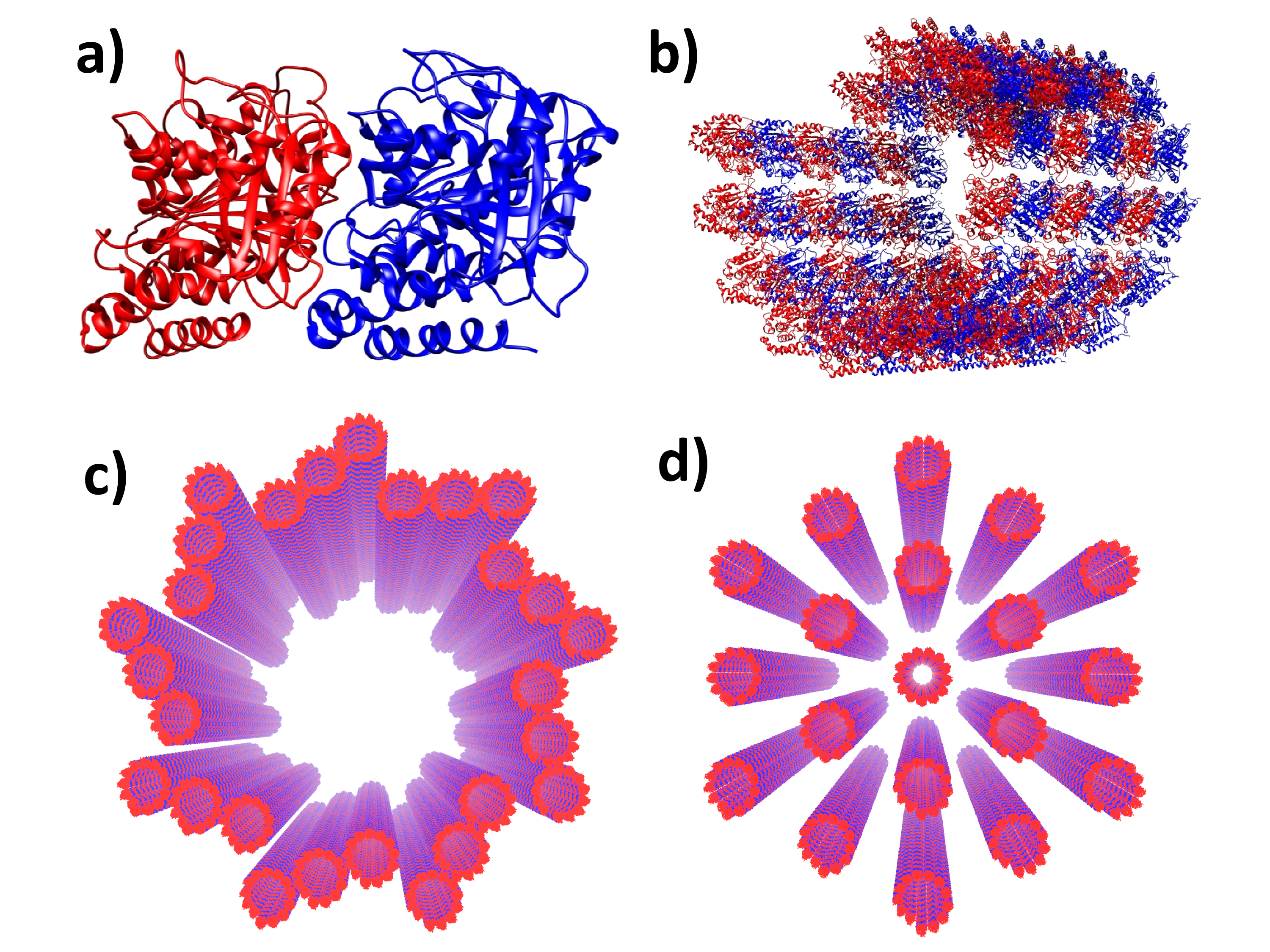}
\caption{\textbf{Hierarchical mega-architectures of tryptophan form in protein aggregates of functional biological significance.} Panels depict a hierarchy of tubulin structures composed of $\alpha$ and $\beta$ tubulin (shown in blue and red
), where panel a) shows an individual tubulin dimer, b) shows a microtubule segment of three dimer-defined spirals, 
c) shows a centriole geometry formed from nine triplets of microtubules, and d) shows a hexagonal bundle of 19 microtubules from a typical mammalian axon. Panels a) and b) were generated with \textit{Chimera}. 
Panels c) and d) were produced using Visual Molecular Dynamics 
on the Argonne Leadership Computing Facility mainframe.}
\label{fig:protogeometries}
\end{figure}

In particular, microtubules (MTs) are macromolecular aggregates of the protein tubulin and represent mesoscale networks of tryptophan residues. 
MTs are spiral-cylindrical tubulin aggregates 
that self-assemble 
to enable cellular reorganization and remodeling for mitosis, differentiation, transport, habitat exploration, and apoptosis~\cite{lee2021myosin}, and they have been found to reorganize structurally under UV irradiation \cite{zaremba_effects_1984, krasylenko2013plant}. In addition, other evolutionarily conserved structures consist of MT architectures, including the centriole, a vortex arrangement generally made of nine ``slats'' of MT triplets (see Fig.~\ref{fig:protogeometries}), which has been the subject of several studies \cite{albrecht-buehler_phagokinetic_1977, albrecht-buehler_rudimentary_1992, albrecht-buehler_cellular_1994, albrecht-buehler_autofluorescence_1997} examining cellular orientation to a light stimulus.

These findings suggest the potential for photophysical and photochemical control of MT dynamics, which have been correlated with the regulation and partitioning of reactive oxygen species (ROS) in living cells \cite{usselman_quantum_2016}. Endogenous, optical ultraweak photon emissions (UPEs) from living organisms are well-documented \cite{cifra_ultra-weak_2014, zapata2021human} in the context of ROS-mediated oxidative stress. Stress-induced ultraviolet UPEs are more prominent during the exponential growth phase of the cellular cycle \cite{quickenden_luminescence_1991, tilbury_luminescence_1992}, implicating them in potential biophotonic signaling along aromatic networks during oxidative metabolism \cite{kurian2017oxidative}. However, the link between cellular metabolic activity, UPEs, and tryptophan network optical dynamics remains far from clear, leaving a critical gap in our knowledge. 

Here, we explore the role of photoexcitation in mesoscale tryptophan networks present in several biological architectures. We show that mega-networks of tryptophan can exhibit a collective optical response in the UV. By analyzing several architectures containing more than $10^5$ tryptophan chromophores---ranging from centrioles to microtubule bundles found in neuronal axons---we predict that strongly superradiant (paired with subradiant) states are often present in their spectra. Combining numerical results and scaling analysis, we determine the strength of the collective response in biological structures of realistic size. The effects of physiological disorder are considered by including fluctuations of the tryptophan excitation energies, demonstrating that the effects of superradiance can survive even at thermal equilibrium. Our predictions are confirmed by our experimental observations of larger quantum yields with increasing tryptophan network size. 

\section{Simulations and Quantum Yield Measurements}

\begin{figure*}[t]
    \centering
    \includegraphics[width=\textwidth]{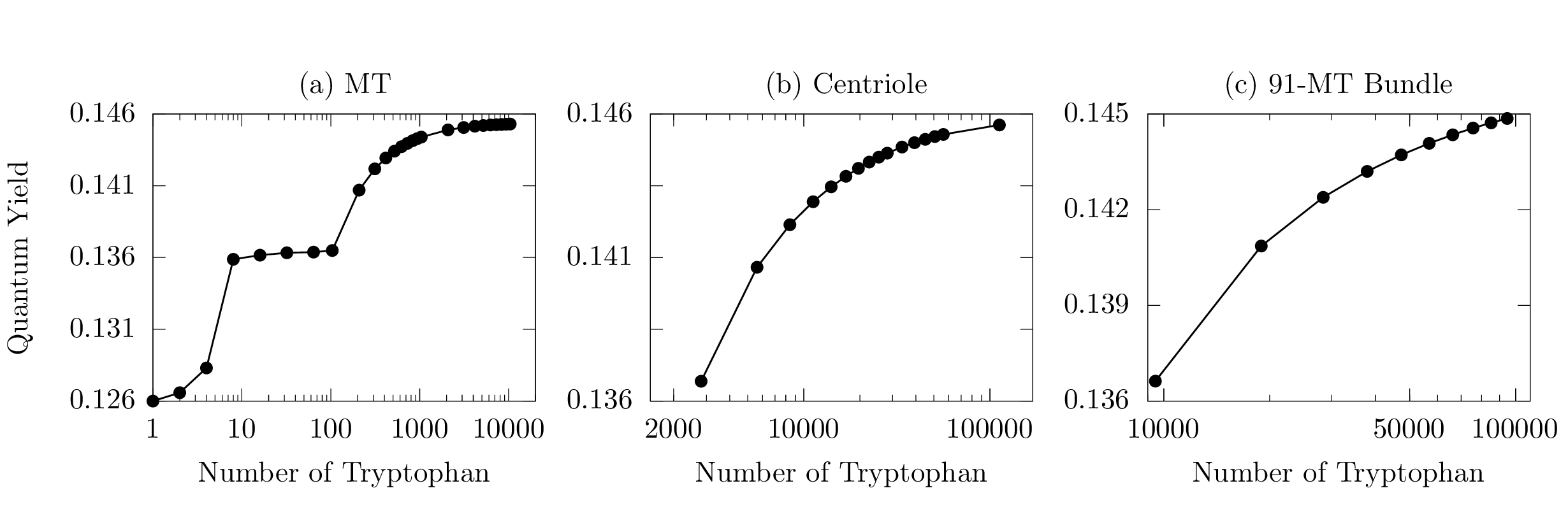} 
    \caption{
   \textbf{Predictions of fluorescence quantum yields from tryptophan networks in protein architectures assuming thermal equilibrium.} The quantum yields (QYs) are plotted as a function of the number of tryptophan
    (Trp) chromophores, where panel (a) shows the QY for a microtubule (MT) (shown in Fig.~1b) with a maximum length of $\sim800\,\text{nm}$, panel (b) shows the QY for a centriole (shown in Fig.~1c) 
    formed by 27 microtubules, each with a maximum length of
    $\sim320\,\text{nm}$, and panel (c) shows the QY for a neuronal bundle formed by 91 microtubules, each with a
    maximum length of $\sim80\,\text{nm}$, arranged in a hexagonal honeycomb (similar to Fig.~1d). 
    }
    \label{fig:MTCentBunQY}
\end{figure*}

We modeled biologically realistic arrangements of tryptophan (Trp) chromophores, beginning with hierarchical aggregates of the protein tubulin as shown in Fig.~\ref{fig:protogeometries} and as described in the Materials and Methods section. We then modeled photoemissive decay channels 
using a well-known radiative non-Hermitian Hamiltonian for open quantum systems (see SI and \cite{mukamelspano1989, mukamelspano1991} for further details). We thus characterized the collective light-matter interaction of Trp mega-networks present in several biologically relevant architectures, solving the complex eigenvalues $E_j -\text{i}\Gamma_j/2$ for each Trp network geometry and determining the radiative decay rates $\Gamma_j$ of the network eigenmodes. Comparing the maximum $\Gamma_j$ with the radiative decay rate $\gamma$ of a single Trp chromophore, we determined the superradiant enhancement factor $\text{max}(\Gamma_j)/\gamma$, thus characterizing each architecture's spectrum by its brightest (\ie, most superradiant) state. Such a model has allowed us to investigate the possibility of whether quantum optical modes may be implicated in photonic coordination of the cytoskeleton and other cellular structures characterized by mesoscale networks of Trp.

The radiative and non-radiative decay processes of the emissive state, in our case the \La \,state of Trp (see SI for further details on distinctions from the \Lb \,state), are quantitatively described by their decay rates $\Gamma$ and $\Gamma_\text{nr}$, respectively~\cite{lakowicz2006principles}. Figure S1 shows the absorption and emission spectra of Trp, tubulin, and microtubules. The absorption-emission Stokes shift is almost identical for tubulin and microtubules (MTs) and significantly smaller than that of Trp. This implies for the protein architectures an overlapping resonance regime between absorption and emission around 300 nm, where the absorptive and emissive transition dipoles are resonant and experimentally indistinguishable. The emissive process is mainly characterized by the observed fluorescence lifetime and the quantum yield. The quantum yield (QY) is defined as the ratio of 
the number of photons emitted to the number of photons absorbed, or equivalently,
\begin{equation}
    \textrm{QY} = \frac{\Gamma}{\Gamma + \Gamma_\text{nr}}\,.
    \label{Eq:QuantumYield}
\end{equation}
We predict the trends of steady-state QYs in the various Trp networks by 
calculating the thermal averages $\langle \Gamma\rangle_\text{th}$ and $\langle \Gamma_\text{nr}\rangle_\text{th}$ of the radiative and 
non-radiative decay rates, respectively, by 
means of the complex eigenvalues of the effective Hamiltonian in Eq.~\ref{Eq_Heff_0E}
(see SI for further details). 

Analyzing various biological architectures of Trp, we  found the emergence of strong superradiant states close to the lowest excitonic state (see SI). The superradiant enhancement increases with the system size until approximately a few times the excitation wavelength, and then it tends toward saturation. The presence of a strong superradiant state close to the lowest-energy state (see Figs.~\ref{fig:axoneme}, \ref{fig:centriole}, and \ref{fig:bigaxon} in the SI) is expected to enhance the QY, since the thermal occupation probability of such a superradiant state---and thus the thermally averaged radiative decay rate---will be enhanced. 

Fig.~\ref{fig:MTCentBunQY} shows the QY predictions for MTs (Fig.~\ref{fig:protogeometries}b), centrioles (Figs.~\ref{fig:protogeometries}c and \ref{fig:centriole}),
and 91-MT bundles (Fig.~\ref{fig:bigaxon}e) of varying lengths. The QY is calculated and displayed in Fig.~\ref{fig:MTCentBunQY} in the form of semi-log plots as
a function of the number of Trp chromophores in the network. 
Here thermalization is assumed at
$k_BT\approx207\, \text{cm}^{-1}$, where Boltzmann's constant is given by $k_B=0.695 \, \text{cm}^{-1}\,\text{K}^{-1}$ and the room-temperature bath by $T=298\,$K.

Fig.~\ref{fig:MTCentBunQY}a shows the case of a MT. Starting with an established experimental value for the QY of Trp \cite{chen1967fluorescence}, the MT QY behavior is divided into 
three regimes. The first one exhibits a rapid, but overall modest increase ($<10\%$) corresponding to the formation of a single tubulin dimer (TuD), containing a total of eight Trp chromophores. The second regime shows near constancy (to 0.1\%) corresponding to the formation of the first MT spiral layer. Each spiral layer contains 13 TuD and a total of 104 Trp chromophores (Fig.~\ref{fig:protogeometries}b). This near-constant regime in the first MT spiral layer can be explained by the fact that the superradiant state is not close to the lowest-energy state for the first spiral, as shown in previous work \cite{celardo2019existence}. The last 
regime for $\text{QY} > 0.136$ shows a familiar sigmoid-like increase and corresponds to the formation of the MT by adding one spiral layer after another, until 100 layers ($\sim800\,$nm) are reached, with a total of $10\,400$ Trp chromophores. QY saturation begins to set in when the MT has reached a length of a few $\lambda$, where 280 nm is the relevant scale set by the wavelength of incident light considered for excitation of the Trp chromophores. Such saturation in the QY is explained by the behavior of the superradiant enhancement factor, which also saturates at this length scale for a variety of structures containing Trp mega-networks (see Figs.~\ref{fig:axoneme}, \ref{fig:centriole}, and \ref{fig:bigaxon} in the SI).

Fig.~\ref{fig:MTCentBunQY}b shows the case of a centriole. The minimum length we consider is a single centriole layer of about 8 nm and containing 2808 Trp chromophores. At 40 layers, we obtain a centriole of approximately 320 nm in length and containing a total of 112\,320 Trp
chromophores. As long as the centriole volume contains between 3\,000 and 20\,000 Trp chromophores, a rapid growth of QY is observed. For larger volumes, the growth of QY slows
down, but the saturation regime is still not fully realized. Fig.~\ref{fig:MTCentBunQY}c shows the case of a 91-MT bundle, with 10 layers and a length of approximately 80 nm. The primary difference in this case is that each layer of the bundle contains 9\,464 Trp chromophores. Similar to the centriole case, the QY increases monotonically as chromophores are added to the network, without realizing saturation even at $10^5$ Trps. 

All three panels in Fig.~\ref{fig:MTCentBunQY} are consistent in showing how thermalization significantly competes with enhancements to the QY from collective effects, without eliminating them, as the superradiance exhibited by these mega-networks in the absence of disorder varies from a few hundreds to several thousands of times the Trp spontaneous emission rate. In panel (c) for example, by increasing the number of chromophores by one order of magnitude, 
from $10^4$ to $10^5$, the increase in thermal QY is only $\sim1\%$. 

These results taken together suggest that equilibrium thermal effects are a primary cause of mitigating such cooperative quantum behaviors, without entirely washing out the associated phenomena. Indeed, we have also considered the effect of structural disorder on the thermal QY by adding time-independent fluctuations of the excitation energies of the Trp chromophores. These fluctuations are typically used to simulate inhomogeneous broadening of the absorption and emission spectra~\cite{spano1989superradiance}. Interestingly we found that the QY is almost unaffected when a disorder strength equal to room-temperature energy ($\sim$200 cm$^{-1}$) is considered, and a QY enhancement is still observable even at 1000 cm$^{-1}$ disorder (see Fig.~\ref{fig:QYW} in the SI). Thus, the QY enhancements presented in Fig.~\ref{fig:MTCentBunQY} are very robust to both thermal environments and structural disorder. 

In order to verify the above theoretical predictions, we performed steady-state fluorescence QY measurements, using the QY of Trp in water \cite{chen1967fluorescence} as a standard. The QYs were determined both for 280-nm excitation, subtracting contributions from other residues, and for 295-nm excitation, where only Trp absorbs (see SI for further details). The QY measurements could only be performed on tubulin and MTs, because of issues with scattered light and sample purity for the larger assemblies, as explained in the SI. Figure S1 shows the steady-state absorption and emission spectra of Trp, tubulin dimers (TuD), and MTs. A scattering background affects the absorption spectrum of MTs and was corrected for as explained in the SI. Table 1 shows consistent results for both excitation wavelengths, with first a decrease of QY from Trp to TuD, then a statistically significant increase by up to almost 70\% for MTs. The decrease from Trp to TuD is small but non-negligible, suggesting that non-radiative processes in the protein are at play. However, this notwithstanding, the significant increases from TuD to MTs are in qualitative agreement with our predictions in Fig.~\ref{fig:MTCentBunQY}a, bearing in mind that our model does not account for additional non-radiative channels due to the formation of large Trp ensembles. It is noteworthy that the increased QY in MTs would imply a decreased non-radiative decay rate, but this is not the case with TuD---rather the contrary. It is therefore an unlikely scenario in the vastly more complex MT case, suggesting that collective radiative processes in these protein assemblies with mega-ensembles of Trp are the primary cause of the significant QY increases observed in MTs.\\

\begin{table}
\begin{center}
 \begin{tabular}[b]{|c | 
   p{25mm} | p{25mm}|}
      \hline
      sample 
      & QY-Trp \newline @ 280 nm (\%) & QY-Trp \newline @ 295 nm (\%) \\ \hline
      microtubules (MT) 
      & $17.6^* \pm 2.1$ & $14.7^* \pm 1.6$  \\ 
      tubulin dimers (TuD) 
      & $10.6 \pm 0.6$ & $10.9 \pm 1.3$  \\ 
    tryptophan (Trp) 
    & $12.4 \pm 1.1$ & $11.4 \pm 1.1$  \\ 
  \hline
    \end{tabular}
        \caption{\textbf{Fluorescence quantum yields from tryptophan networks in protein architectures.} 
    Summary of experimental measurements obtained from steady-state spectroscopy of tryptophan, tubulin dimers, and microtubules in BRB80 aqueous buffer solution (see Fig.~\ref{Fig:AbsFlu_TrpTubMT} for complete spectra). Fluorescence quantum yield (QY) is determined for excitation at 280 nm 
    and 295 nm (see Materials and Methods for details about the procedure). 
    Note the statistically significant increases in the QY from tubulin to microtubules, in qualitative agreement with Fig.~\ref{fig:MTCentBunQY}a and consistent with what one would expect in the presence of superradiance. The $^*$ indicates an average of upper and lower limit values for microtubules, which have been corrected for the scattering background.
    }
    \label{tab:QY-main}
        \end{center}
    \end{table}

\section{Discussion}

The fluorescence response from multi-tryptophan proteins, with different lifetime components conventionally associated to different classes of tryptophan based on the heterogeneity of local environments \cite{khan_steady-state_1997, Schenkl05, leonard_functional_2009}, becomes even more complicated when considering mega-networks of tryptophan residues formed by their biological architectures. In this work, we have simulated collective photoexcitonic properties of such extremely large Trp networks in protein structures ranging from individual tubulin dimers and microtubule segments to microtubule super-architectures such as the centriole and neuronal bundles (Fig.~\ref{fig:protogeometries}).

Even though the coupling between Trp transition dipoles is relatively weak ($\sim60$ cm$^{-1}$) compared to room-temperature energy ($\sim200$ cm$^{-1}$), the presence of long-range couplings between Trp chromophores can greatly enhance the robustness of the network~\cite{celardo2019existence}. Moreover, a counter-intuitive consequence of cooperativity  is the fact that the robustness of a system to disorder can increase with the system size. This effect, known as cooperative robustness in the literature, has been investigated theoretically in  paradigmatic models~\cite{celardo2014cooperative,celardo2014cooperative2,chavez2019real}. 
In this work we have shown that mega-networks of Trp in protein architectures can exhibit cooperative robustness, as shown in the bottom panel of Fig.~\ref{fig:centriole}.
The origin of this effect can be qualitatively explained as a very large decay width strongly coupling the system with the electromagnetic field. Such strong coupling protects the system from disorder, which must become comparable to the coupling in magnitude to suppress superradiance. 

On the other hand, our findings also reveal the fundamental challenges of coherent quantum optical information transfer at ambient temperatures in the presence of static and/or dynamical disorder. Significant disorder can effectively quench collective superradiant effects, even though our fluorescence quantum yield measurements of tryptophan, tubulin, and microtubules in aqueous buffer solution suggest that even in thermal equilibrium such effects survive. Certainly, more robust models are needed to account for exciton-phonon couplings in deformations of the protein scaffold \cite{faraji_electrodynamic_2022}, as well as for optical pumping of mechanical modes in non-equilibrium structural organization and assembly \cite{nardecchia2018out, zhang_quantum_2019, azizi_examining_2022}. .

Microtubules are crucial to cytoskeletal regulation and form complex bundles in neuronal tissue. Our studies of axonal microtubule bundles (Figs.~\ref{fig:protogeometries}d, \ref{fig:MTCentBunQY}c, \ref{fig:bigaxon}, and SI) may have implications for both neuroscience and quantum optics research. 
Confining a superradiant optical mode to one dimension in a waveguide has been proposed to extend emitter interactions to extremely long range \cite{solano2017super}, raising the tantalizing possibility that axons might serve as such waveguides between giant superradiant emitters in the brain. 
Microtubule bundles in axons or those associated with the centrosome complex may satisfy a particular combination of criteria necessary to exhibit these ultra-long-range couplings, which are currently being exploited for state-of-the-art chiral nanofiber communications systems \cite{solano2017super, kim2018super,solano2017optical}.

\section{Conclusions}
Although the roles of Trp as a metabolic precursor 
and fluorescent reporter have been studied in depth, the implications of large Trp architectures for photophysical control of biosystems remain largely unexplored. Trp chromophores have been identified for their unique role in UV light sensing in the UVR8 plant photoreceptor \cite{wu2012structural}, which is believed to be the first UV light perception system discovered to use a network of Trp chromophores as a funnel to enhance its quantum efficiency \cite{li_leap_2020}. This utilization of a network of intrinsic amino acids for light sensing marks a significant departure from other photoreceptors, which rely on a separate cofactor (such as flavin adenine dinucleotide in cryptochrome) or pigment (such as chlorophyll in photosynthesis or retinal in rhodopsin) to enable light detection and harvesting. Recent observations of UV light-harvesting from Trp networks in microtubules \cite{kalra2022electronic} and of the Trp network as a photoreduction mediator in cryptochrome \cite{lin2018circadian} are consistent with an emerging picture of extended protein scaffolds that harness the symmetries of hierarchical Trp networks to promote biological function.

Past studies elucidated the physical plausibility of superradiant effects in individual microtubule geometries of varying lengths \cite{celardo2019existence}, and in this work we extend these findings to study tryptophan networks of vastly increased scale, revealing how collective and cooperative quantum effects might manifest in cytoskeletal networks and other protein aggregates associated with diverse cellular structures and organelles. We have also analyzed the collective quantum optical response of microtubule bundles present in neuronal axons, where photons from brain metabolic activity could be absorbed rapidly via superradiant states for ultrafast information transfer.

Our work highlights essential features of tryptophan chromophore networks in large aggregates of proteins forming biomolecular super-architectures such as the centriole (Figs.~\ref{fig:protogeometries}c, \ref{fig:MTCentBunQY}b, \ref{fig:centriole}), axoneme (Fig.~\ref{fig:axoneme}), and microtubule bundles in neurons (Figs.~\ref{fig:protogeometries}d, \ref{fig:MTCentBunQY}c, \ref{fig:bigaxon}). Specifically, by analyzing the coupling with the electromagnetic field of mega-networks of tryptophan present in these biologically relevant architectures, we find the emergence of collective quantum optical effects, namely superradiant and subradiant eigenmodes. Our analysis has been done using a radiative Hamiltonian (see Eq.~\ref{Eq_Heff_0E} in the SI) in the single-excitation limit, which is reasonable given the biological milieu of ultraweak photon emissions. The presence of collective superradiant eigenmodes in such a wide variety of biological complexes---and their observed manifestation in increasing QYs for larger hierarchies of proteins---suggests that this collective ultraviolet response would be exploitable \textit{in vivo}. 

Exceptionally bright superradiant states in these biocomplexes may facilitate the absorption and energy transfer of UV photoexcitations in an intensely oxidative environment, where electronically excited molecular species emit light quanta in this wavelength regime. In this manner, superradiant states promoting enhanced QYs for large biological architectures may serve a photoprotective role in pathological conditions such as Alzheimer's disease and related dementias, since an enhanced QY implies that a greater portion of the photonic energy absorbed by certain protein aggregates is re-emitted rather than assimilated by those complexes. Such collective and cooperative mechanisms for photoprotection have not been fully explored, even in the case of the black-brown pigment eumelanin, which consists of a mixture of two indole monomers that aggregate to form oligomers of different lengths and geometries. A recent study of eumelanin \cite{ilina_photoprotection_2022, chergui_funneling_2022} demonstrated ultrafast energy transport over large distances despite the significant structural and chemical inhomogeneity of the sample, raising the question of whether mega-networks of indole from tryptophan and neuromelanin can aid in ``internal'' UV energy downconversion and funnelling in the brain. Similarly, the UV superradiant response in mega-networks of tryptophan could also augment artificial light-harvesting devices to extend and enhance the spectral band of absorption beyond the visible range.

Our work thus presents numerous possibilities for superradiance- (and subradiance-)enabled metabolic regulation, communication, and control in and between cells (see Table \ref{TableScales}), and with external agents that interact with the cytoskeleton at various stages of cellular growth and replication \cite{oliva2022effect}. Combined with experimental measurements of fluorescence QY in tubulin architectures, our simulations presented in this work demonstrate that collective and cooperative UV excitations in Trp mega-networks support robust quantum states in proteins with observable consequences even under thermal equilibrium conditions.

\section{Materials and Methods}
\subsection*{Protein Structural Models}
We created computer models of these realistic biological geometries using atomic coordinates of proteins downloaded from the Protein Data Bank (PDB). We extracted the Trp coordinates (positions and orientations) from each PDB file to create tables of transition dipole moment coordinates as in Ref.~\cite{celardo2019existence}, choosing the well-known \La \, peak excitation at $280\,\text{nm}$ as our transition dipole moment of interest \cite{Schenkl05}.

We used the Trp transition dipole coordinates obtained for each structure to define matrix elements of the radiative Hamiltonian given in Eq.~\ref{Hmuk} (see SI and Fig.~\ref{matrixelements} for further details). The complex eigenvalues of Eq.~\ref{Eq_Heff_0E} in the SI contain information on the emission spectra $\{E_j\}$ and linewidths $\{\Gamma_j\}$  in the single-excitation limit \cite{celardo2019existence}). We simulated these resonances by diagonalizing the matrix in Eq.~\ref{Eq_Heff_0E} for each Trp arrangement (see SI). 
These spectra allowed us to predict optical enhancements due to collective quantum optical interactions in the Trp networks found in a variety of prototypical cellular structures, organelles, and appendages.

Tubulin (Fig.~\ref{fig:protogeometries}a) was modeled using PDB entry 1JFF, and tubulin dimers were assembled into a virtual microtubule (Fig.~\ref{fig:protogeometries}b)  according to the protocol given in Appendix A of Ref.~\cite{celardo2019existence}: aligning the (would-be) outer microtubule surface with the $y$ axis (by rotating it -55.38$^\circ$ in the $yz$ plane transverse to the microtubule longitudinal $x$ axis),  optimizing the tubulin orientation (\ie, rotating each dimer 11.7$^\circ$ around the $\beta$-tubulin Trp346 C$_{\delta 2}$ atom in the $yz$ plane and then translating the dimer 0.3 nm in the $z$ direction), before translating each dimer 11.2 nm in the $y$ direction, and successively rotating it by multiples of $-27.69^\circ$ in the $yz$ plane (about the origin around the $x$ axis) while successively shifting each dimer by multiples of 0.9 nm in the $x$ direction.

Larger microtubule architectures were built virtually from arrangements of individual microtubules that were constructed as described above. Microtubule bundles (from model axons as shown in Fig.~\ref{fig:protogeometries}d and Fig.~\ref{fig:bigaxon}) were prepared in hexagonal arrangements by placing adjacent microtubules $50\,$nm apart center-to-center, reflecting the mean separation between adjacent tau-mediated microtubules in experimentally observed neuronal axons \cite{chen1992projection}. The model centriole as shown in Fig.~\ref{fig:protogeometries}c and Fig.~\ref{fig:centriole} was created from an initial triplet of microtubules centered $100\,\text{nm}$ from the origin along the $y$ axis---with the microtubules in this triplet centered at $(x,y,z)$ coordinates (0, 87, -22.5167), (0, 100, 0), and (0, 113, 22.5167) in nm---before each triplet was rotated in increments of $40^\circ$ around the origin in the $yz$ plane. The idealized (1JFF) model axoneme was constructed as ten pairs of microtubules, with each microtubule pair spaced $26\,\text{nm}$ apart center-to-center. One microtubule pair is centered at the origin, and the remaining nine pairs are spaced evenly (40$^\circ$ apart) and centered at a distance of $98\,\text{nm}$ from the origin as shown in Fig.~\ref{fig:axoneme}.

\subsection*{Prediction of Quantum Yield}
The fluorescence quantum yield (QY) is the ratio of the number of emitted photons (per time and volume unit) relative to the number of absorbed ones. Equivalently, the QY can be defined in terms of the radiative decay rate $\Gamma$ and the non-radiative decay rate $\Gamma_{\text{nr}}$: QY$= \Gamma/(\Gamma+\Gamma_{\text{nr}})$, 
where $\Gamma_{\text{nr}}$ represents all the physical and chemical processes involved in the interaction between the chromophore network and the surrounding protein(s) or solvent. Typically we can write, $\Gamma_{\text{nr}} = \Gamma_{\text{IC}} + \Gamma_{\text{ISC}} + 
\Gamma_{\text{react}}$, where $\Gamma_{\text{IC}}$ is the internal conversion rate constant; 
$\Gamma_{\text{ISC}}$ is the intersystem crossing rate constant, and $\Gamma_{\text{react}}$ is 
the rate constant due to quenching or photochemical reactions.

The effective Hamiltonian $\Heff$ from Eq.~\ref{Eq_Heff_0E} in the SI is the starting point for our QY predictions. To consider the effects of non-radiative processes in our model, 
we replace the diagonal part of $\Heff$ with a new decay rate $\gamma' = \gamma + \gamma_{\text{nr}}$. Here $\gamma_{\text{nr}}$ represents the decay rate of a single Trp due to 
non-radiative processes. Then, the new eigenvalues of $\Heff$ are given by $\mc{E}_j'=
E_j'-\ii\,\Gamma_j'$ where $\Gamma_j' = \Gamma_j + \Gamma_{\text{nr},j}$.
The QY is a dimensionless quantity that takes values between 0 and 1. When $\Gamma \gg 
\Gamma_\text{nr}$, QY $\rightarrow 1$, but if the excited state depopulation is dominated by quenching processes, external conversion, or intersystem crossing, $\Gamma_\text{nr} \gg \Gamma$ and QY $\rightarrow 0$. 

For the particular case of the Trp chromophore, the radiative decay rate $\gamma=0.00273\,
\text{cm}^{-1}$ corresponds to a radiative lifetime $\tau=1.9$ ns~\cite{celardo2019existence}. At room temperature its quantum yield in water is estimated to be QY $\approx\, 0.13$~\cite{chen1967fluorescence}, although in different proteins the Trp QY has been observed to vary from about one-tenth this value to nearly a factor of three times it~\cite{callis_quantitative_2004}. Using Eq.~\ref{Eq:QuantumYield} with the replacements $\Gamma \rightarrow \gamma$ and $\Gamma_\text{nr} \rightarrow \gamma_\text{nr}$ allows us to calculate the Trp non-radiative decay rate in water as $\gamma_\text{nr} \approx 0.0183$ cm$^{-1}$.  

\textit{Thermal average.} Consider $P(t)$ as the probability that an excitation is found in the chromophore network at time $t$, while $1-P(t)$ would be the probability that the excitation has left the network. Let us denote as $P_k(t)$ the probability that the chromophore system is decribed by the eigenstate $\ket{\ER_k}$ at time $t$. Assuming thermal equilibrium, 
$P_k(t)=P(t)\exp(-E_k/k_BT)/Z$. Here $k_B=0.695$ cm$^{-1}$
K$^{-1}$ stands for the Boltzmann constant, $T$ is the temperature, and $Z=\sum_{j=1}^N
\exp(-E_j/k_BT)$ is the partition function. 

Due to the non-Hermitian nature of $\Heff$, the probability $P(t)=\sum_{k=1}^N P_k(t)$ at 
thermal equilibrium is not conserved ($\dot{P}(t)\neq 0$). Instead it satisfies the following master equation, $\dot{P}(t)=-\average{\Gamma}_\text{th}P(t)$, where the thermal 
average for the decay rate is given by $\langle\Gamma\rangle_\text{th}=\sum_{j=1}^N\Gamma_j
\exp(-E_j/k_BT)/Z$. Because $\gamma_j$ for each Trp is assumed equal to all others, and our model neglects the formation of additional non-radiative channels with increasing Trp network size, we consider $\average{\Gamma_{\text{nr}}}_\text{th} \rightarrow \gamma_{\text{nr}}$ and arrive at the following definition for the QY at thermal equilibrium:
\begin{equation}
    \average{\text{QY}}_\text{th} = \frac{\average{\Gamma}_\text{th}}{
    \average{\Gamma}_\text{th} + \average{\Gamma_{\text{nr}}}_\text{th}}\,.
\end{equation}

\subsection*{Measurement of Quantum Yield}
Steady-state fluorescence spectroscopy was performed with a Shimadzu RF-5301PC spectrofluorophotometer. Conventional absorption spectra were taken with a Shimadzu UV-3600 UV-Vis spectrophotometer. For the measurements we used UV-grade glass cuvettes with a pathlength of 1 cm.

\textit{Samples.} Tubulin protein in the form of $\alpha-\beta$ tubulin heterodimers and pre-formed microtubules (taxol-stabilized and lyophilized) extracted from porcine brain were purchased from Cytoskeleton, Inc. The microtubules exhibit an average length of 2 µm. For tubulin and tryptophan (Sigma-Aldrich) solutions we used a self-prepared BRB80 buffer (80 mM PIPES pH 6.9, 2 mM MgCl$_2$, and 0.5 mM EGTA, pH 6.9). Microtubules were stabilized in solution by adding 20 µM taxol (Cytoskeleton, Inc.) to the BRB80 buffer. The proteins are delivered as powders in 1 mg vials (TuD) and 0.5 mg vials (MTs). However, the vials contain a bit more than 1 mg and 0.5 mg, respectively, according to the manufacturer. Moreover, around 5 mg of sucrose plus 1 mg Ficoll is added to the vials. Hence, it was not possible for us to prepare solutions with exact concentrations. We used $\sim$0.33 mg/ml tubulin protein and microtubules for steady-state spectroscopy. Tyrosine was measured in ultrapure water solution and cysteine in 1 mM HCl. The absorption background by the solvents was subtracted.

\textit{Absorption and emission spectra.} Steady-state absorption and emission spectra of microtubules (MTs), tubulin protein ($\alpha-\beta$ tubulin heterodimers, TuD) and tryptophan (Trp) in physiological BRB80 buffer are shown in Figs.~\ref{Fig:AbsFlu_TrpTubMT} and \ref{Sxx-Rayleighcorr}. They reveal a first absorption band with a maximum around 280 nm for all three systems. The MTs solution exhibits a strong scattering background (Fig.~\ref{Sxx-Rayleighcorr}d), which needed to be subtracted. Assuming Rayleigh-like scattering ($\propto \lambda^{-4}$), we fitted the background from 307 to 800 nm. The extrapolated curve for wavelengths $\lambda <$ 307 nm was subtracted from the raw spectrum. The corrected spectrum agrees qualitatively well with the TuD spectrum for wavelengths above $\sim$270 nm. The upper limit for the MT quantum yield is determined from the corrected spectrum (by subtracting the fit, resulting in a lower sample absorbance $a_s$ in Eq.~\ref{fluo_ref}), and the lower limit for the MT quantum yield is determined from the raw data without correction. 

The normalized fluorescence spectra upon excitation at 280 nm show that Trp fluorescence has its maximum at 355 nm, while for TuD and MTs the absorption-emission Stokes shifts are significantly smaller than for Trp, and almost identical for both (fluorescence maxima at $\sim$327 nm). We attribute it to a reduced chromophore-solvent interaction due to the protein environment. The full-width half-maximum (FWHM) intensity values of TuD and MTs are experimentally identical ($\sim$5.5·10$^3$ cm$^{-1}$) and are broader than for Trp (4.9·10$^3$ cm$^{-1}$). 

TuD consist of 8 Trp, 35 tyrosine (Tyr), and 20 or 21 cysteine amino acids that can form up to 10 cystine (Cys) residues linked in pairs by disulfide bonds, even though most of these are too far apart to form such a bridge. In order to roughly estimate the molar absorption coefficient of the protein at $\sim$280 nm, we simulated the absorption spectrum of TuD by adding the contributions of the Trp (5.6·10$^3$ M$^{-1}$ cm$^{-1}$, 49\% contribution), Tyr (1.3·10$^3$ M$^{-1}$ cm$^{-1}$, 50\% contribution), and Cys residues (125 M$^{-1}$ cm$^{-1}$, 1\% contribution), i.e., the number of the respective residues multiplied with their respective molar absorption coefficients at $\sim$280 nm~\cite{pace_proteinsci_1995}. This yields a molar absorption coefficient of tubulin of
92·10$^3$ M$^{-1}$ cm$^{-1}$. The reconstructed spectrum is in good agreement with the experimental one down to 270 nm (see Fig.~\ref{Sxx-Rayleighcorr}c). However, the experimental spectrum is slightly broadened, probably by inhomogeneous contributions. The other amino acids from the protein’s backbone have negligible contributions ($\ll$1\%) to the molar absorption coefficient in this range. The deviations below 270 nm are presumably due to contributions of the more than 800 remaining amino acids of the protein backbone including, \eg, phenylalanine (43 residues) or alanine (80 residues).

Among these amino acids, Trp and Tyr exhibit the strongest fluorescence quantum yields (Trp 13\% and Tyr 14\% in H$_{2}$O~\cite{lakowicz2006principles}). The fluorescence maximum of TuD is blueshifted by 2.2·10$^3$ cm$^{-1}$ with respect to Trp and redshifted by 2.9·10$^3$ cm$^{-1}$ with respect to Tyr (see Fig.~\ref{Sxx-Rayleighcorr}b). Hence, reconstruction of the TuD fluorescence spectrum by adding the spectra of the two components, even with optimized weights, fails. 

\textit{Observation of fluorescence quantum yields.} Fluorescence quantum yields (QYs) were determined according to the standard reference formula \cite{lakowicz2006principles},
\begin{equation}
\text{QY}_s=(F_s a_r n_s^2)/(F_r a_s n_r^2 )\, \text{QY}_r.
\label{fluo_ref}
\end{equation}
The subscripts $s$ and $r$ represent the examined sample and the reference, respectively. $F$ is the integrated fluorescence area, and $n$ is the refractive index of the respective solution. The absorption factors $a$ are determined by $a=1-10^{-A}$, with $A$ the optical density at the absorption wavelength of the absorption band. The QY of Trp in H$_2$O ($\sim$13\%) was used as reference~\cite{chen1967fluorescence, brouwer_standards_2011}. We adjusted the concentrations of the samples to approximately equal optical densities in the absorption maxima of the various samples. In order to minimize statistical errors, the spectra of five freshly prepared solutions for each sample were averaged.

The QY of Trp in the BRB80 buffer (12.4 ± 1.1\%) does not deviate significantly from its value in H$_2$O. Tubulin exhibits a QY of 6.8 ± 0.4\%, which is reduced with respect to Trp. This indicates the contribution of other chromophores to the absorption band on the one hand and the dominant Trp fluorescence on the other hand. Lacking knowledge of the exact contribution of scattering to the optical density of the MTs solution at 280 nm, we can only give a range or average for the quantum yield of MTs. By subtracting the extrapolated fit from the raw data, 
we estimate 12.0 ± 1.0\% as an upper limit of the MT QY. Its lower limit is given by employing the MT absorption without any corrections of scattering and yields 10.3 ± 0.8\%. 

In order to determine the QY generated by the Trp residues in the TuD and in MTs, we weighted the absorption spectrum by its contribution from Trp, which is 49\% at 280 nm. Therefore, the QY for TuD grows to 10.6 ± 0.6\% and for MTs we get 19.5 ± 2.8\% for its upper limit and 15.7 ± 1.3\% for its lower limit (Table 1 of the main text and Table~\ref{tab:QY-SI}). TuD still exhibits a reduced QY with respect to Trp, but the MT QY from Trp is clearly enhanced.

In order to minimize error due to overlapping absorption by other residues, we also evaluated the QY at 295 nm (instead of the absorption maximum at 280 nm) excitation, where the absorption of  amino acids other than Trp is negligible. The values are also given in Table~\ref{tab:QY-SI}, and they confirm the trends obtained for 280 nm excitation.

\bibliography{scibib}

\bibliographystyle{Science}

\begin{scilastnote}
\item The project was supported by The Guy Foundation Family Trust. This research used resources of the Argonne Leadership Computing Facility, which is a DOE Office of Science User Facility supported under Contract DE-AC02-06CH11357. This research used molecular graphics and analyses performed with \textit{UCSF Chimera}, developed by the Resource for Biocomputing, Visualization, and Informatics at the University of California, San Francisco, with support from NIH P41-GM103311. We would also like to acknowledge insightful discussions with Patrik Callis.
\end{scilastnote}

\clearpage

\renewcommand{\thefigure}{S\arabic{figure}}
\setcounter{figure}{0} 

\renewcommand{\thetable}{S\arabic{table}}
\setcounter{table}{0} 

\renewcommand{\thesection}{S\arabic{section}}
\setcounter{section}{0} 

\clearpage
\section*{Supplementary Material}

\section{Non-Hermitian Open Quantum Systems}

\begin{table}[bh]
    \centering
    \caption{
    Evolution of Hermitian $vs$ non-Hermitian systems}
 \resizebox{\textwidth}{!}{   \begin{tabular}{ll}
         &  \\
        Hermitian case & non-Hermitian case      \\ \hline
                       &                         \\
        \multicolumn{2}{l}{\textbullet\, \underline{Schrödinger equation}} \\[0.2cm] 
        $\ii\frac{d}{dt}\ket{\psi(t)}=\hat{H}\ket{\psi(t)}$ \quad ($\hat{H}^\dagger=\hat{H}$) & 
        $\ii\frac{d}{dt}\ket{\psi(t)}=\Heff\ket{\psi(t)} \quad (\Heff^\dagger\neq\Heff)$  \\
        & $\Heff = \hat{H}_0 - \frac{\ii}{2}\hat{\Gamma}$ \quad ($\hat{H}_0^\dagger=\hat{H}_0$ 
        and $\hat{\Gamma}$ is a real and symmetric matrix) \\
        \multicolumn{2}{l}{\textbullet\, \underline{Evolution operator}} \\[0.2cm] 
        $\ket{\psi(t)} = \U(t)\ket{\psi(0)}$ & $\ket{\psi(t)} = \UU(t)\ket{\psi(0)}$ \\
        with the unitary $\U(t)=\exp(-\ii\hat{H}t)$ \ie $\,\U\U^\dagger=\U^\dagger\U=\one$  & 
        with the non-unitary $\UU(t)=\exp(-\ii\Heff t)$ \ie $\,\UU\UU^\dagger=\UU^\dagger\UU=\one\ee^{-\Gamma t}$ \\
                                             &                                       \\
        \multicolumn{2}{l}{\textbullet\, \underline{Inner product}} \\[0.2cm] 
        $\braket{\psi(t)}{\psi(t)} = 1$      & $\braket{\psi(t)}{\psi(t)} = \sum_m C_m^{\textrm{R}}C_m^{\textrm{L}}\ee^{-\Gamma_m t}$ \\
                                             &                                       \\
        $\bra{\psi(t)} = \bra{\psi(0)}\U^\dagger = \bra{\psi(0)}\ee^{\ii\hat{H}t}$  & 
        $\bra{\psi(t)} = \bra{\psi(0)}\UU^\dagger(t)$ \\
                                             &            \\
                                             & with $C_m^{\textrm{R}}=\braket{\psi(0)}{\mc{E}_m^{\textrm{R}}}$ 
                                             and $C_m^{\textrm{L}}=\braket{\mc{E}_m^\textrm{L}}{\psi(0)}$ \\
                                             &                                       \\
        \multicolumn{2}{l}{\textbullet\, \underline{Evolution of an ensemble $\hat{\varrho}$}} \\[0.2cm] 
        $\hat{\varrho}(t) = \U(t)\hat{\varrho}(0)\U^\dagger(t)$ & $\hat{\varrho}(t) = \UU(t)\hat{\varrho}(0)\UU^\dagger(t)$ \\
        with $\hat{\varrho}(0)=\sum_\alpha w_\alpha\ketbra{\psi(0)}{\psi(0)}$ and $\sum_\alpha w_\alpha = 1$ &
        with $\hat{\varrho}(0)=\sum_\alpha w_\alpha\ketbra{\psi(0)}{\psi(0)}$ and $\sum_\alpha w_\alpha = 1$ \\
        properties: $\forall\,t\,\,$, (i) $\tr(\hat{\varrho})=1$, (ii) $\hat{\varrho}^\dagger=\hat{\varrho}$ and (iii) $\hat{\varrho} > 0$ &
        properties: for $t\neq 0$, (i) $\tr(\hat{\varrho})\neq 1$, (ii) $\hat{\varrho}^\dagger=\hat{\varrho}$ and (iii) $\hat{\varrho} > 0$ \\
                                             &                                       \\
        \multicolumn{2}{l}{\textbullet\, \underline{Liouville equation}} \\[0.2cm] 
        $\frac{d}{dt}\hat{\varrho}(t)=-\ii[\hat{H}, \hat{\varrho}]$ & 
        $\frac{d}{dt}\hat{\varrho}(t)=-\ii[\hat{H}_0, \hat{\varrho}] - 
        \frac{\ii}{2}(\hat{\Gamma}\hat{\varrho} + \hat{\varrho}\hat{\Gamma}) $ \\ 
                                             &                                       \\
        \hline
    \end{tabular} }
    \label{tab:H_vs_nH}
\end{table}

The dynamics of the chromophore network are governed by the time-dependent Schrödinger equation
\begin{equation}
    \frac{d \ket{\psi(t)}}{dt} = -\frac{\ii}{\hbar} \Heff\,\ket{\psi(t)}\,,
\end{equation}
which results in a non-unitary evolution given
the non-Hermitian nature of the Hamiltonian $\Heff$ from Eq.~\ref{Eq_Heff_0E} in the SI. If we denote as $\{\ket{\ER_j}\}$ the right 
eigenvectors of $\Heff$, $\Heff\ket{\ER_j} = \mc{E}_j\ket{\ER_j}$ with the
complex eigenvalue $\mc{E}_j=E_j-\ii\Gamma_j/2$. As the set of right eigenvectors form an
orthonormal basis for the Hilbert space, then the state of the system for $t>0$ can be 
written as a linear combination of those states: $\ket{\psi(t)}=\sum_j
\CL_j\,\ee^{-\ii\mc{E}_j t/\hbar}\ket{\ER_j}$, where $\CL_j=\braket{\EL_j}{\psi(0)}$,
$\ket{\psi(0)}$ is the initial state, and $\bra{\EL_j}$ is the left eigenvector corresponding 
to $\ket{\ER_j}$. Since the Hamiltonian is symmetric, 
the left eigenvectors $\bra{\EL_j}$ are defined as the transpose of the right eigenvectors $\ket{\ER_j}$.

Since the standard inner product used in Hermitian systems, where $\bra{\psi(t)}$ is 
precisely defined as the conjugate transpose of the respective $\ket{\psi(t)}$, is not applicable here, it is necessary to introduce an 
alternative definition, commonly known as the Euclidean inner product (or sometimes the ``c-product''~\cite{moiseyev2011non}). The time evolution of the initial $\bra{\psi_0}$ is then given by $\bra{\psi(t)}=\sum_j
\bra{\EL_j}\CR_j\,e^{\ii\mc{E}_j^* t}$, with $\CR_j=\braket{\psi_0}{\ER_j}$ such 
that $\braket{\psi(t)}{\psi(t)}=\sum_{j} \CR_j\CL_j\, e^{-\Gamma_j t}$. This definition 
allow us to calculate the time average of any observable we wish to study. In Table~\ref{tab:H_vs_nH} of the SI we list some of the main differences between 
Hermitian and non-Hermitian systems.

\begin{figure*}
\centering
\includegraphics[width=0.8\textwidth]{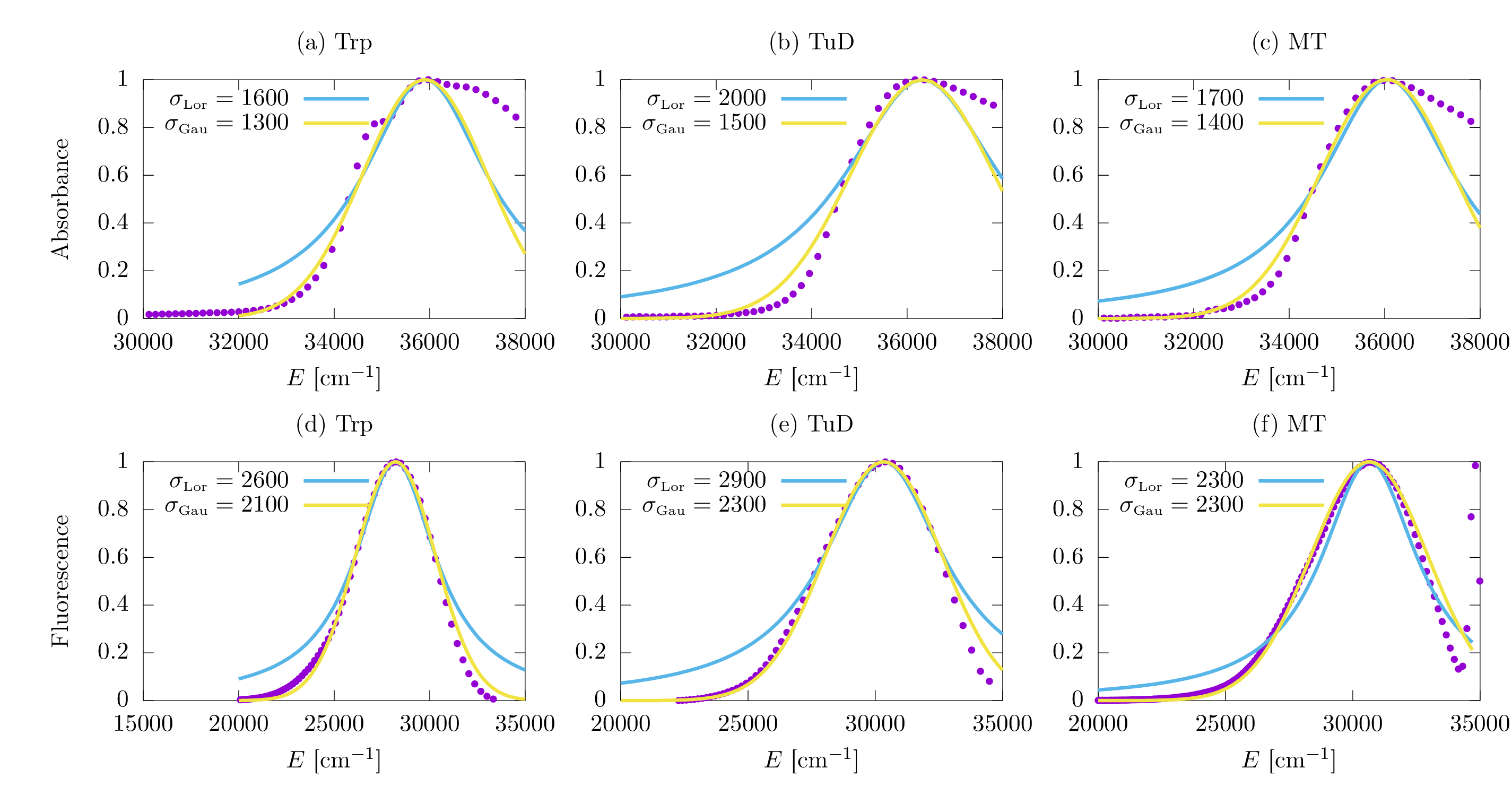}
\caption{\textbf{Absorption and fluorescence spectra for tryptophan (Trp), tubulin dimers (TuD), and microtubules (MT) in aqueous solution.} Comparison between experimental data (purple dots) and numerical estimates (solid curves). Two line shapes, 
Lorentzian (blue) and Gaussian (yellow), are considered for the numerical estimates. The homogeneous 
broadening is introduced by the parameter $\sigma$. 
}
\label{Fig:AbsFlu_TrpTubMT}
\end{figure*}

\begin{table}
\begin{center}
\resizebox{\textwidth}{!}{%
\begin{tabular}[b]{|c|p{15mm}| p{15mm}| p{25mm}|p{25mm} | p{23mm}| p{23mm}|}
     \hline
      sample & abs \newline max (nm) & fluo \newline max (nm) & FWHM fluo \newline @ 280 nm ($\times 10^3 \, \text{cm}^{-1}$) & FWHM fluo \newline @ 295 nm ($\times 10^3 \, \text{cm}^{-1}$) & QY-Trp \newline @ 280 nm (\%) & QY-Trp \newline @ 295 nm (\%) \\ \hline
      microtubules (MT) & $277 \pm 0.5$ & $327 \pm 0.5$ & $5.5 \pm 0.1$ & $5.4 \pm 0.1$ & $17.6^* \pm 2.1$ & $14.7^* \pm 1.6$  \\ 
      tubulin dimers (TuD) & $277 \pm 0.5$ & $328 \pm 0.5$ & $5.6 \pm 0.1$ & $5.6 \pm 0.1$ & $10.6 \pm 0.6$ & $10.9 \pm 1.3$  \\ 
    tryptophan (Trp) & $278 \pm 0.5$ & $355 \pm 0.5$ & $4.9 \pm 0.1$ & $5.0 \pm 0.1$ & $12.4 \pm 1.1$ & $11.4 \pm 1.1$  \\ 
     \hline
    \end{tabular}}
        \caption{\textbf{Steady-state spectroscopy data from tryptophan networks in protein architectures.} 
    Summary of experimental measurements obtained from steady-state spectroscopy of tryptophan, tubulin dimers, and microtubules in BRB80 aqueous buffer solution (see Fig.~\ref{Fig:AbsFlu_TrpTubMT} for complete spectra). Abbreviations: Absorption maximum (abs max), fluorescence maximum (fluo max), fluorescence quantum yield for Trp contributions at 280 nm (QY-Trp @ 280 nm), fluorescence quantum yield at 295 nm (QY-Trp @ 295 nm), and fluorescence bandwidths at full-width half-maximum for excitation at 280 nm (FWHM fluo @ 280 nm) and 295 nm (FWHM fluo @ 295 nm), respectively. Fluorescence QY is determined from excitation either at 280 nm, where contributions from other amino acids have been subtracted, or at 295 nm, where only tryptophan absorbs and the contributions from other residues can be neglected. Note the statistically significant increases in the QY from tubulin to microtubules, in qualitative agreement with Fig.~\ref{fig:MTCentBunQY}a and consistent with what one would expect in the presence of superradiance. The $^*$ indicates an average of upper and lower limit values for microtubules, which have been corrected for the scattering background.
    }
    \label{tab:QY-SI}
        \end{center}
    \end{table}

\section{Superradiance phenomena and model of tryptophan quantum optical networks} \label{Sec:La_model}
It is well known in quantum optics that organized networks of quantum two-level systems can exhibit a phenomenon called \textit{superradiance}, also known as superfluorescence. In 1954, Dicke \cite{dicke1954coherence} predicted this behavior, which involves a collection of identical light emitters spontaneously emitting intense coherent radiation. Since then, superradiance has been observed in various physical systems such as molecular aggregates \cite{de1990dephasing,fidder1990superradiant}, cold atoms \cite{araujo2016superradiance}, diamond nanocrystals \cite{bradac2017room}, semiconductor quantum dot ensembles \cite{brandes2005coherent,scheibner2007superradiance}, and more recently in both nanocrystal superlattices \cite{raino2018superfluorescence,philbin2021room,cherniukh2021perovskite} and hybrid perovskite thin films \cite{findik2021high}.

Typically, the probability density that a single excited chromophore emits a photon is exponentially distributed and is characterized by a decay rate $\gamma$. If there are $N$ emitters, all in the excited state, superradiance theory predicts cooperative emission with $\sim N^2$ times higher peak intensity than that of a single emitter. If the incident radiation is so weak that only one excitation is present, single-excitation superradiance \cite{scully2009super} can result. 
The hallmark of this cooperative quantum effect, where a single excitation is 
coherently shared by $N$ emitters,
is a decay rate that scales proportionally with $N\gamma$.

The analysis of the coupling of the tryptophan network with the electromagnetic field in the single-excitation limit can be effectively made with a radiative Hamiltonian widely used in quantum optics, which allows treatment of systems whose size is much larger than the absorbed wavelength~\cite{scully2009super}. Superradiant states can assist photon absorption at specific frequencies and enhance excitation transfer to or along other aggregates due to supertransfer processes~\cite{celardo2019existence,gulli}. Moreover, the presence of superradiant (short-lived, bright) states always comes together with the presence of subradiant (long-lived, dark) states, which can be used to store the absorbed excitation energy. 

The tryptophan (Trp) networks we consider are modeled as an ensemble of $N$ two-level systems, each characterized by a transition dipole ($\vec{\mu}_n$). 
The interaction of a network of two-level systems with the electromagnetic field
is described by the effective
Hamiltonian~\cite{mukamelspano1989,mukamelspano1991,robinakkermans2008}
\begin{equation}
  \hat{H}_{\text{eff}} = \hat{H}_0 + \hat{\Delta} - \frac{\text{i}}{2}\, \hat{\Gamma}\, , \label{Hmuk}
\end{equation}
where $\hat{H}_0$ represents the sum of the excitation energies of each Trp chromophore, and
$\hat{\Delta}$ and $-\text{i}\, \hat{\Gamma}/2$ represent the coupling matrices between chromophores induced by interaction with the electromagnetic field. The non-Hermiticity in the term $\text{i}\, \hat{\Gamma}/2$ arises from the fact that the photoexcitation can be lost to the field leading to non-unitary evolution (\ie, spontaneous emission). 

The effective Hamiltonian in Eq.~\ref{Hmuk} has been widely used to model
light-matter interactions in the limit of a single excitation, which is reasonable given the biological milieu of ultraweak photon emissions. 
We consider the primary contribution to the spectra due to the collective interactions between all pairs of Trp choromophores contained in these aggregates \cite{celardo2019existence}, such that $\Heff$ may be expanded as

\vspace{-0.4cm}
\begin{equation}
   \label{Eq_Heff_0E}
    \!\!\Heff \!=\! \sum_{n=1}^{N} \!\left( \hbar \omega_0 - \ii\frac{\gamma}{2} \right) \! \Prj{n} \, + \!\!
    \sum_{{m,\,n = 1;\\
            m\neq n}}^{N} \!\!\!\left( \Omega_{m n} - \ii\frac{\Upsilon_{\!m n}}{2} \right)\!\ketbra{m}{n},
\end{equation}
where $E_0 = \hbar\omega_0$ is the excitation energy and $\gamma = 4\mu^2 k_0^3/3$ is the spontaneous emission rate of the Trp transition dipole moment $\mu = |\vec{\mu}|$. The angular wavenumber is $k_0 = 2\pi/\lambda$, where $\lambda$ is the wavelength of the incident light required to bring the Trp chromophore from the ground state to the excited state. Matrix elements $\Omega_{m n}$ 
and $\Upsilon_{m n}$ represent the couplings between the $N$ enumerated Trp transition dipoles in the
network, induced by interaction with the electromagnetic field~\cite{celardo2019existence,grad1988radiative}:

\begin{eqnarray}
    \Omega_{m,n} &= \frac{3 \gamma}{4}\left\{ 
    -\left[ \hat{\mu}_m\cdot\hat{\mu}_n -  (\hat{\mu}_m\cdot\hat{r}_{mn})(\hat{\mu}_n\cdot\hat{r}_{mn}) \right]
    \frac{\cos(k_0 r_{mn})}{k_0 r_{mn}} + \right. \nonumber \\
     \hspace{0.0cm}[ \hat{\mu}_m\cdot\hat{\mu}_n &- \left.3(\hat{\mu}_m\cdot\hat{r}_{mn})(\hat{\mu}_n\cdot\hat{r}_{mn}) ]
    \left[  \frac{\sin(k_0 r_{mn})}{(k_0 r_{mn})^2} + \frac{\cos(k_0 r_{mn})}{(k_0 r_{mn})^3} \right]
    \right\}
\end{eqnarray}
\begin{eqnarray}
    \Upsilon_{m,n} &= \frac{3 \gamma}{2}\left\{ 
    \,\left[ \hat{\mu}_m\cdot\hat{\mu}_n -  (\hat{\mu}_m\cdot\hat{r}_{mn})(\hat{\mu}_n\cdot\hat{r}_{mn}) \right]
    \frac{\sin(k_0 r_{mn})}{k_0 r_{mn}} + \right. \nonumber \\
     \hspace{0.0cm}[ \hat{\mu}_m\cdot\hat{\mu}_n &- \left.3(\hat{\mu}_m\cdot\hat{r}_{mn})(\hat{\mu}_n\cdot\hat{r}_{mn}) ]
    \left[  \frac{\cos(k_0 r_{mn})}{(k_0 r_{mn})^2} - \frac{\sin(k_0 r_{mn})}{(k_0 r_{mn})^3} \right]
    \right\}.
\end{eqnarray}
Here $\hat{\mu}_m = \vec{\mu}_m/\mu_m$ is the unit dipole moment of the $m$th Trp, and $\hat{r}_{mn}=
(\vec{r}_n-\vec{r}_m)/r_{mn}$, where $r_{mn}=|\vec{r}_n-\vec{r}_m|$ stands for the distance between the $m$th and $n$th Trps. 

The eigenvalues of the complex symmetric matrix in Eq.~\ref{Eq_Heff_0E} can be decomposed into real and imaginary parts, which respectively designate the excitation energies $\{E_j\}$ and decay rates $\{\Gamma_j\}$ of the fluorescent ensemble the matrix describes. Unlike Hermitian operators used to represent systems of bound states, Eq.~\ref{Eq_Heff_0E} describes a set of scattering \textit{resonances}. The values for the physical parameters considered in our analysis are~\cite{celardo2019existence}:
\begin{itemize}
\item $\lambda = 280\,\text{nm}$ ($E_0 = 35714$ cm$^{-1}$) as the Trp peak excitation wavelength (energy),
\item $ k_0 = 2 \pi E_0 = 2.24 \times 10^{-3}$ \AA$^{-1}$ as the angular wavenumber,  
\item $\mu=6$ Debye as the strength of the transition dipole between the ground state and the first excited state, with $\mu^2 \approx 181224 \mbox{  \AA}^3 \mbox{  cm}^{-1}$ (for the conversion, see~\cite{celardo2019existence} and for further information on transition dipole states, see below), and 
\item $ \gamma= 4 \mu^2 k_0^3 / 3 = 2.73 \times 10^{-3} \,\text{cm}^{-1}$, where $ \gamma/\hbar $ is the radiative decay rate of a single Trp molecule, corresponding to the radiative lifetime $\tau \approx 1.9 \, \text{ns}$ (for the conversion, see~\cite{celardo2019existence}).
\end{itemize}
 
Microtubule structures are comprised mainly of the tubulin dimer (Fig.~\ref{fig:protogeometries}), which gives the structures their characteristic spiral-cylindrical shape. Formed from a pair of subunits denoted $\alpha$ and $\beta$, tubulin proteins house a diversity of chromophores, such as tryptophan, tyrosine, and phenylalanine residues. The internal structure of these chromophores, in the form of their aromatic moieties, confer upon them quantum optical properties such as their well-characterized transition dipole moments. Thus, these chromophores can be considered like small two- or three-level systems that absorb light of a certain wavelength and emit it, generally at a different wavelength. 

When ordered networks of these chromophores interact with the electromagnetic field, their transition dipoles may synchronize coherently to give rise to superradiance. To characterize this behavior in protein systems as large as the functional axonemes (Fig.~\ref{fig:axoneme}), centrioles (Fig.~\ref{fig:centriole}), and neuronal microtubule bundles (Fig.~\ref{fig:bigaxon}) described in this work, appropriate simplification of the quantum degrees of freedom must be performed. First, our model only includes tryptophan (Trp) chromophores in each protein structure, because its primary electronic transition dipole moment is considerably larger than that of all other amino acids. Second, assuming that the intensity of the incident light is sufficiently weak, we only consider the limit of the single-excitation manifold. Third, we do not consider any higher electronic transition dipole moments nor the vibronic transitions (0-1, 0-2, etc.) of Trp, for reasons described in the next section. This allows us to describe the interaction between the chromophore network and the electromagnetic field by means of the non-Hermitian effective Hamiltonian in Eq.~\ref{Eq_Heff_0E}, similar to the tight-binding Hamiltonians typically used in solid-state physics and for photosynthetic light-harvesting complexes~\cite{gulli}.

\section{Clarification between the \La~and \Lb~transition dipoles}

\begin{figure*}[hb]
    \centering
    \includegraphics[width=0.62\textwidth,trim={0 0 0 0},clip]{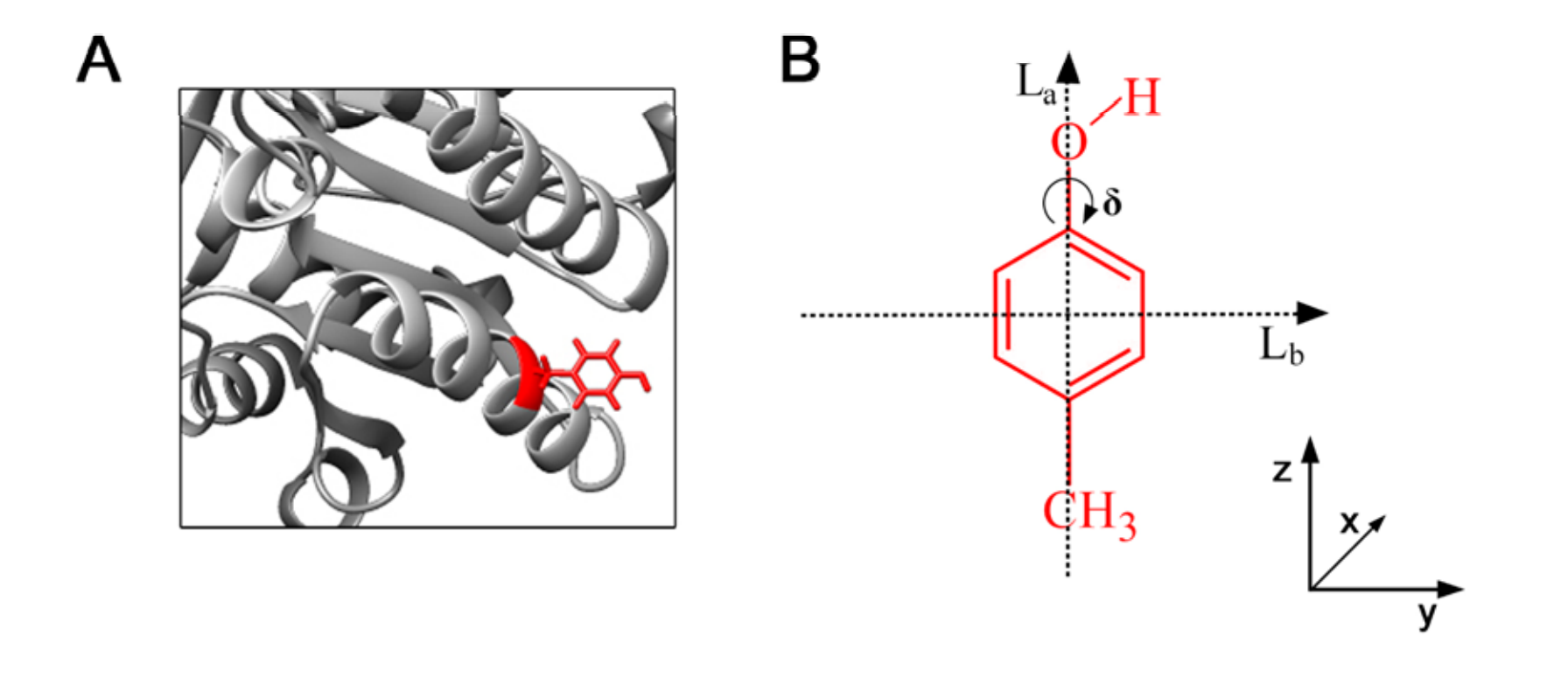}
    \caption{\label{fig:TyrosineDipole} Panel \textbf{A} shows protein backbone and Tyr amino acid residue, alongside panel \textbf{B} showing chemical structure formula of Tyr residue with directions of \La \, and \Lb \, transition dipole moments. Reproduced from \cite{fornander2014uv}.}
\end{figure*}

The \La~and \Lb~transition dipoles of cyclic aromatics take their nomenclature from the geometric orientation of the dipole moments with respect to the aromatic plane, where the dipole vectors representing the transitions are centered in the plane such that the \La~vector overlaps with the \textit{atoms}, whereas the \Lb~vector overlaps with the \textit{bonds} (Fig.~\ref{fig:TyrosineDipole}). While this geometric definition of \La~and \Lb~is unambiguous in more benzene-like aromatics such as tyrosine (Tyr) and phenylalanine (Phe), the meaning is obscured in indoles such as tryptophan (Trp) where the orthogonal (perpendicular) dipole moments \La~and \Lb~are angled so that neither one clearly corresponds to a respective ``atom'' or ``bond'' axis.

To avoid confusion associated with the aforementioned geometrical ambiguity in Trp, we consider two orthogonal excited states of tryptophan that consist of a larger transition dipole moment that is about twice the strength of the associated smaller one. For the purpose of modeling absorption, we concern ourselves with the 0--0 (purely electronic) transitions rather than the vibronic ones:  The 0--0 transition of the larger transition dipole occurs at $\sim 300$ nm (according to Valeur and Weber \cite{valeur1977resolution}), and the 0--0 transition of the smaller one occurs at $\sim 290$ nm. One should take notice from Valeur and Weber~\cite{valeur1977resolution} in their Figure 6 that the size of the \Lb~peak is about half that of the \La~value at that wavelength ($\sim 290$ nm), suggesting that the \Lb~transition dipole strength is about half that of \La ~\cite{callis19977,lombardi1999solvatochromic}.

Use of the terms \La~and \Lb~to describe these transitions in Trp persists in the literature, despite the ambiguity of the subscripts in Trp. To clarify this, we may generally consider the \La~electronic transition to be more broadened than the \Lb~transition in polar solvents because of the larger \La~transition dipole moment that induces a lowering of the \La~energy~\cite{lombardi1999solvatochromic}. To include vibronic transitions (0--1, 0--2, \etc), the \La~spectrum will have portions that are both lower and higher in energy than that of the \Lb~spectrum because of this significant line-broadening effect.

On a related note, it is necessary to distinguish the transition spectra of isolated indole from the spectra of the Trp residue itself. For example, Callis [Figure 6 in Ref.~\cite{callis19977}] validated calculations for indole \La~and \Lb~spectra in a polar solvent matrix, by comparing these data to results from Valeur and Weber~\cite{valeur1977resolution}---but only after re-scaling and translating his calculated values to account for the spectral differences between Trp and indole proper. Thus, Figures 5 and 6 from Valeur and Weber~\cite{valeur1977resolution} provide actual representations of tryptophan spectra; in particular, the 0--0 transition of \La~(not the peak) is red-shifted with respect to that of \Lb.

\section{Absorption and emission spectra}\label{Sec:AbsFluo}

\begin{figure*}
\centering
\includegraphics[width=0.6\textwidth]{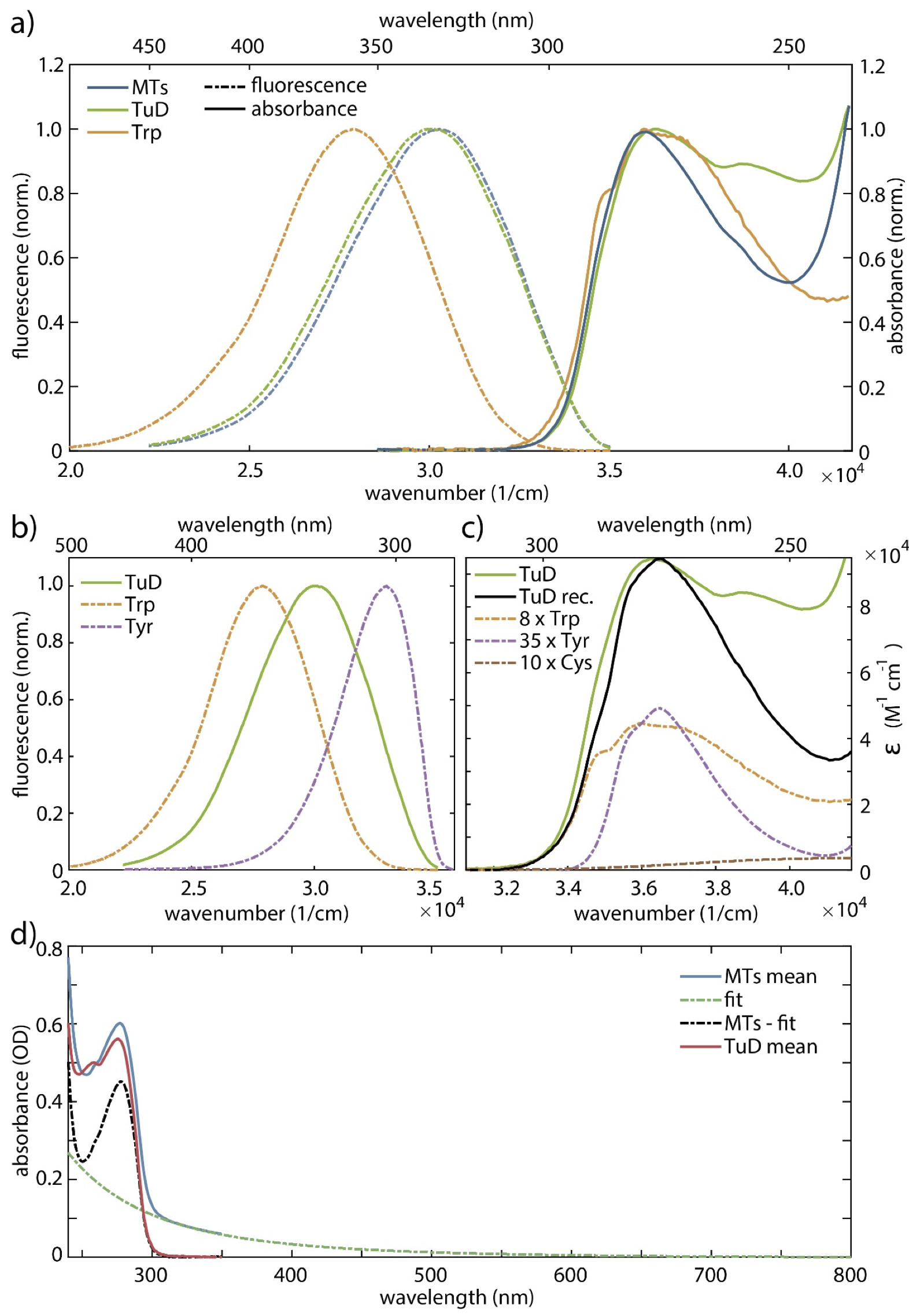}
\caption{\textbf{Steady-state absorption and fluorescence spectra of whole proteins and reconstruction from primary constituents in aqueous solution.} a) Normalized fluorescence spectra (dash-dotted) of MTs (blue), TuD (green) and Trp (orange). Absorption spectra (solid) are normalized to the peak absorption maxima. All spectra are corrected by subtracting background due to solvent absorption scattering. b) Normalized fluorescence spectra of TuD (solid green) and its primary fluorescent constituents Trp (dash-dotted orange) and Tyr (dash-dotted violet).  c) Reconstruction of the molar absorption coefficient at 280 nm of TuD (solid black) by adding different contributions of Trp (8 residues, dash-dotted orange), Tyr (35 residues, dash-dotted violet) and Cys (10 residues, dash-dotted brown). The measured absorption spectrum of TuD from a) is added for comparison (solid green). d) Fit of the scattering background of the MTs solution according to a Rayleigh-like model $\propto \lambda^{-4}$ (green dashed). The MTs mean spectrum (blue) was used for determination of the lower limit of the QY, and the background-subtracted spectrum MTs-fit (black dashed) was used for the upper limit of the QY. The mean tubulin spectrum (red) is plotted for comparison.
}
\label{Sxx-Rayleighcorr}
\end{figure*}

The absorption spectrum indicates the amount of incident electromagnetic radiation absorbed 
by the chromophores in the network, between a range of energies or frequencies. Following the 
work of Renger and Marcus~\cite{RengMarc02}, we can define the linear absorption spectrum as the real part of a Fourier-Laplace transform of the dipole-dipole correlation 
function, which can be expressed as a function of the energy as
\begin{equation}
    \mathcal{A}(E) = A \sum_j \Gamma_j D_j(E)
    \label{Eq_Absorption02}
\end{equation}
where $\Gamma_j$ is the decay rate corresponding to the $j$th eigenstate of the $\Heff$, 
$D_j(E)$ is known as the \textit{lineshape} function, and $A$ is a normalization factor. Under the 
assumption of Markovian behavior, there is insufficient time for the excitation of vibrational quanta, and a 
Lorentzian lineshape function centered at 
$ E_j$ is obtained:
\begin{equation}
    D_j(E) = \frac{\sigma}{(E-E_j)^2+\sigma^2}\,.
    \label{Eq_Absorption_1Exc}
\end{equation}
On the other hand, in the strong coupling regime a Gaussian lineshape is expected:
\begin{equation}
    D_j(E) = \exp[-(E-E_j)^2/2\sigma^2].
\end{equation}
The parameter $\sigma$ measured in units of cm$^{-1}$ is associated with the homogeneous broadening of the on-site chromophore energies. These analytical lineshapes are compared with experimental steady-state absorption and fluorescence spectra for Trp, tubulin dimers, and microtubules in Fig.~\ref{Fig:AbsFlu_TrpTubMT}. Additional details on the experimental steady-state spectra are catalogued in Table~\ref{tab:QY-SI} and in Fig.~\ref{Sxx-Rayleighcorr}.

Conversely, the fluorescence spectrum indicates the amount of electromagnetic radiation emitted by the chromophores in the network. The expression for the fluorescence emission intensity is obtained by multiplying
each lineshape by the corresponding Boltzmann factor
\begin{equation}
    I(E) = A'\sum_j \frac{\ee^{-E_j/k_BT}}
    {Z}\Gamma_j D_j(E),
    \label{Eq_Fluorescnce_1Exc}
\end{equation}
where the partition function $Z=\sum_l\exp[-(E_l)/k_BT]$, and $A'$ is a distinct normalization factor. Here the temperature $T$ is measured in Kelvin ($K$), and $k_B$ is Boltzmann's constant.

\begin{table}[bh]
\caption{
Synopsis of superradiant and subradiant features of tryptophan networks (of size $N$) in biological structures, where $\tau_j = (2\pi c\,{\Gamma_j})^{-1}$ and MT = microtubule. (Analytical fits for superradiance have been used for the first two entries, hence subradiance data is not available.)} 
\begin{tabular*}{0.75\textwidth}{lc|ccc}
$\displaystyle\mathrm{Protein\; Structure,}\atop\displaystyle\mathrm{Length\; in\; nm}$ & $\displaystyle\frac{{\!\max}(\Gamma_{\! j})}{N\gamma}$ & $\tau_\text{super}$ (ps) & $\displaystyle{\!\min}\Bigg{(}\frac{\Gamma_{\! j}}{\gamma}\Bigg{)}$ & $\tau_\text{sub}$ (s) \\ 
\hline 
91-MT Axon, 320 (fit) & 0.012  & 0.428 & N/A & N/A \\
61-MT Axon, 320 (fit) & 0.016 & 0.479 & N/A & N/A \\
Centriole, 400 & 0.028 & 0.495 & $4.6\times 10^{-8}$ & 0.042 \\
61-MT Axon, 224 & 0.020  & 0.547 & $3.6\times 10^{-10}$ & 5.4 \\
37-MT Axon, 320 & 0.026 & 0.602 & $2.3\times 10^{-10}$ & 8.5\\
91-MT Axon, 152 & 0.017  & 0.636 & $2.6 \times 10^{-10}$ & 7.5 \\

Axoneme (1JFF), 320 & 0.031 & 0.754 & $ 2.8 \times 10^{-10}$ & 6.9\\
19-MT Axon, 320 & 0.032 & 0.769 &
$9.9\times 10^{-10}$ & 2.0 \\
7-MT Axon, 640 & 0.039 & 0.856 & $1.4\times 10^{-10}$ & 13.9 \\
7-MT Axon, 320 & 0.071 & 0.941 & $2.8\times 10^{-9}$ & 0.69 \\
Axoneme (6U42), 320 & 0.010 & 2.64 & $1.0\times 10^{-8}$ & 0.19 \\
1 Microtubule, 320 & 0.119 & 3.93 & $7.8\times 10^{-8}$ & 0.025\\
\end{tabular*}
\label{TableScales}
\end{table}

\section{Simulations of axoneme superradiance}
\begin{figure*}[tbhp]
\centering
\includegraphics[width=.53\linewidth,trim={0 0 0 30},clip]{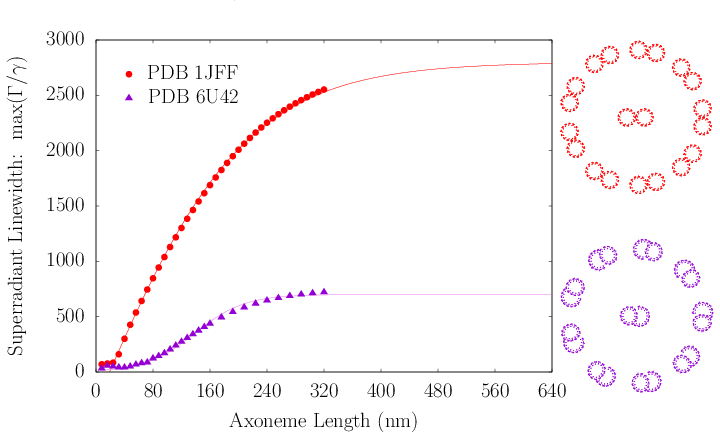}
\includegraphics[width=.462\linewidth,trim={0 0 0 30},clip]{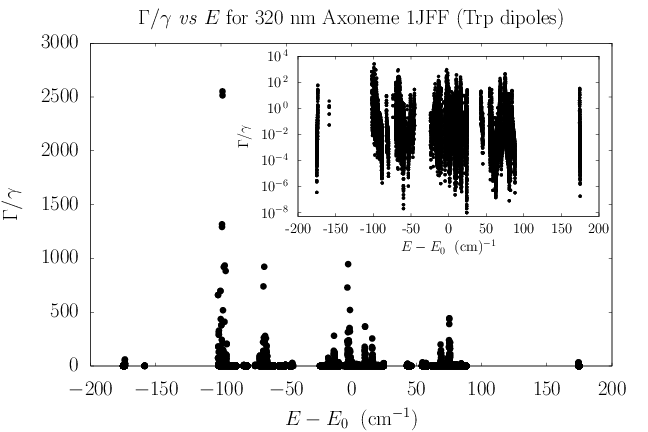} 
\caption{\textbf{Ciliary axoneme structures exhibit superradiance, but less so than centrioles and neuronal microtubule bundles of comparable length.} Left panel shows superradiance data points $\max(\Gamma/\gamma)$ calculated from numerical diagonalization of the radiative Hamiltonian in Eq.~\ref{Eq_Heff_0E} for a model axoneme (generated using tubulin dimers modeled from PDB entry 1JFF) in red, approximated by the curve $f_\text{1JFF}^\text{\,axo}(\ell)=\frac{\lambda n_\text{D}}{\ell_0}[(n_\text{S}-2)\tanh(\ell/2n_\text{S}\ell_0)-1]$, where $\ell$ is the axoneme length along the longitudinal axis (in nm), $\ell_0 = 8\,\text{nm}$ denotes the longitudinal length of a single tubulin spiral, $\lambda=280\,\text{nm}$ is the excitation wavelength, $n_\text{D}=8$ is the number of Trp dipoles per tubulin dimer, and $n_\text{S}=13$ is the number of dimers per tubulin spiral. Likewise, the $\max(\Gamma/\gamma)$ data points for a more realistic axoneme (constructed using a ciliary doublet from PDB entry 6U42) are shown in violet, fit by the curve $f_\text{6U42}^\text{\,axo}(\ell) =\frac{\lambda (n_\text{S}-3)}{\ell_0}[\tanh(3\ell/2n_\text{S}\ell_0-2) +1]$. 
Axoneme cross-sections are shown between the two panels as arrays of point dipoles representing the Trp transitions ($L_\text{a}$) in the colors red (1JFF) and violet (6U42). The right panel shows the spectrum ($\Gamma/\gamma$ \textit{vs} $E - E_0$, where $E_0$ is the excitation maximum of Trp) of the $320\,\text{nm}$-long model 1JFF axoneme containing 83200 Trp dipoles, plotted on linear and semi-log (inset) scales.}
\label{fig:axoneme}
\end{figure*}

The \textit{axoneme} is the microtubule-based structural core of the flagellum or cilium of a eukaryotic cell \cite{ishikawa2017axoneme}. It typically contains nine microtubule doublets surrounding a central one (Fig.~\ref{fig:axoneme}). For comparison, we solved the spectrum of an axoneme modeled as an idealized array of microtubule pairs, again generated from the tubulin protein crystal structure (PDB entry 1JFF)~\cite{lowe2001refined}. We also considered a more realistic model axoneme based on a ciliary microtubule doublet obtained by cryo-electron microscopy (PDB entry 6U42) \cite{ma2019structure}. Even though the primary function of the axoneme is locomotive and mechanical, both the idealized axoneme and ciliary doublet simulations predicted significant superradiant enhancements in the values of $\max(\Gamma_j/\gamma)$, as shown in the left panel of Fig.~\ref{fig:axoneme}. The right panel of Fig.~\ref{fig:axoneme} shows the energy spectrum of the 1JFF axoneme spread over a range of $E_0 \pm 200\,\text{cm}^{-1}$ around the Trp peak excitation $E_0$.

\section{Simulations of centriole superradiance}

\begin{figure*}
\centering
\includegraphics[width=0.5\linewidth,trim={0 0 0 40},clip]
{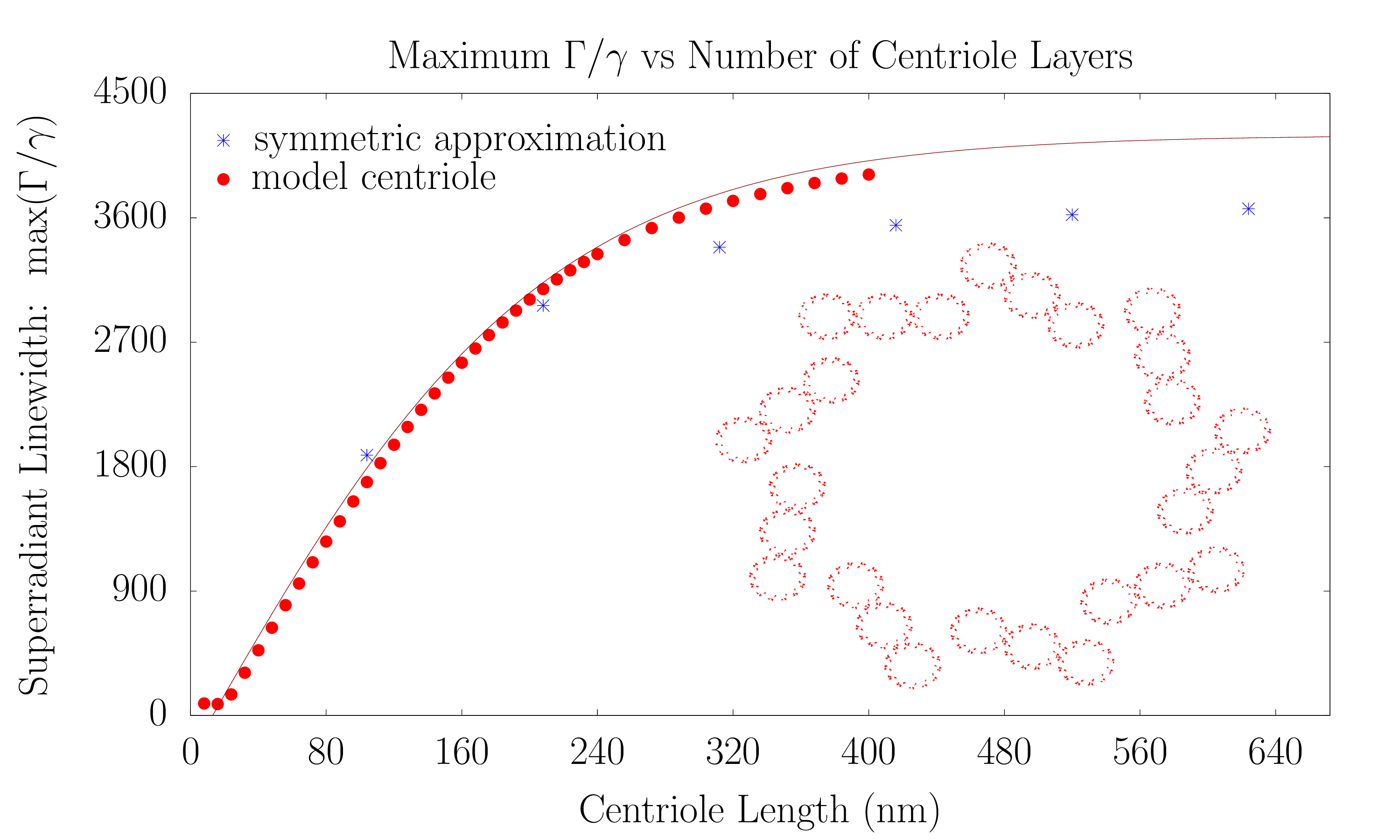}
\includegraphics[width=
0.5\linewidth,trim={0 0 0 30},clip]{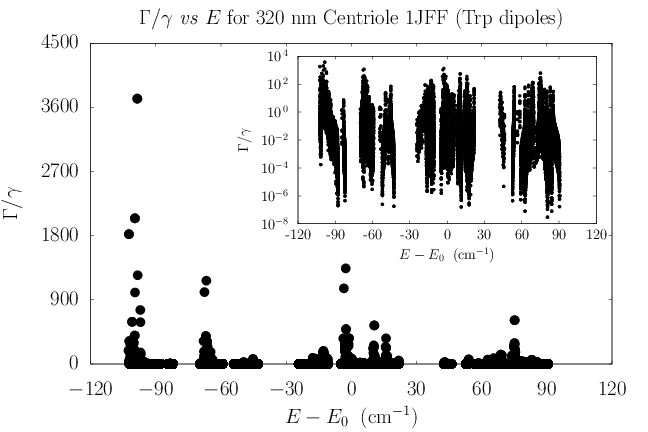}
\includegraphics[width=
0.5\linewidth,trim={0 0 0 30},clip]{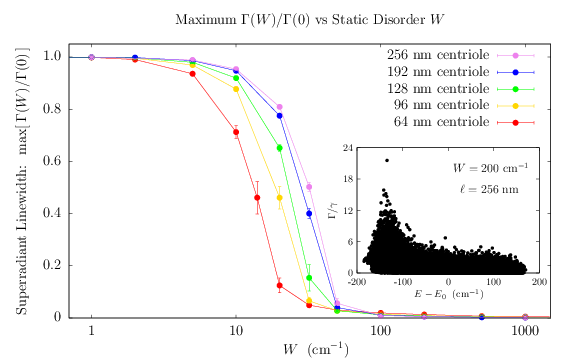}
\caption{\textbf{Prediction of superradiant states in the centriole and the robustness of superradiance to on-site disorder in centriolar tryptophan architectures.} Top panel shows red superradiance data points $\max(\Gamma_j/\gamma)$ calculated from numerical diagonalization of the radiative Hamiltonian in Eq.~\ref{Eq_Heff_0E} for a model centriole Trp architecture,  
approximated by the curve $f_\text{1JFF}^\text{\,cent} (\ell)=\frac{\lambda n_\text{D}}{\ell_0}[2n_\text{D}\tanh(\ell/2n_\text{S}\ell_0)-1]$, where $\ell$ denotes the centriole length along its longitudinal axis (in nm), $\ell_0 = 8\,\text{nm}$ denotes the longitudinal length of a single tubulin spiral, $\lambda=280\,\text{nm}$ is the excitation wavelength, $n_\text{D}=8$ is the number of Trp transition dipoles (\La) per tubulin dimer, and $n_\text{S}=13$ is the number of dimers per tubulin spiral. Blue stars represent the approximate predicted values 
for the brightest state given by Eq.~\ref{eq:approxcentriole}. 
The top panel inset shows the centriole cross-section as point dipoles representing the Trp transition states. 
Center panel shows the spectrum 
of a $320$ $\text{nm}$-long centriole containing 112320 Trp dipoles, plotted on linear and semi-log (inset) scales. Bottom panel shows superradiance data points 
with static disorder for model centrioles of lengths $\{\ell\}$. 
The bottom panel inset shows the spectrum for a $256$ $\text{nm}$-long centriole containing 89856 Trp dipoles at $W = 200\,\text{cm}^{-1}$ (\textit{i.e.}, commensurate with $k_\text{B} T$ at a temperature $T\approx 288\,\text{K}$).
\label{fig:centriole}}
\end{figure*}

Dynamic organization of the MT network is coordinated in vertebrate cells from the centrosome \cite{nigg2011centrosome}, 
which is comprised of two perpendicular 
centrioles \cite{winey2014centriole} in most eukaryotes. 
The \textit{centriole} (Fig.~\ref{fig:protogeometries}c) is one of the largest (protein-based) structures of the cell, exhibiting a cartwheel-like arrangement of microtubules that can be as large as $250\,\text{nm}$ in diameter and up to $500\,\text{nm}$ in length in vertebrates~\cite{winey2014centriole}. It is a pinwheel-shaped, barrel-like organelle that coordinates cellular orientation and division processes. 

We modeled a prototype centriole of increasing length as an array of nine microtubule triplets, where each simulated microtubule was generated from the structure of tubulin PDB entry 1JFF~\cite{lowe2001refined} following past work~\cite{celardo2019existence}. We numerically solved the spectrum of each centriolar Trp arrangement to predict the enhancement factors $\max(\Gamma_j/\gamma)$ shown in the top panel of Fig.~\ref{fig:centriole}. We found the maximum superradiant enhancement increased with growing length until saturating at $\max(\Gamma_j/\gamma)\approx 4000$, in the realistic length scale of vertebrate centrioles. This large superradiant enhancement is identified with a state in the lowest band of excitonic eigenstates of the Hamiltonian of~Eq.~\ref{Hmuk}, as displayed in the center panel of Fig.~\ref{fig:centriole} for a 320-nm centriole Trp architecture, where the variation in the energies spans about $E_0 \pm 100\,\text{cm}^{-1}$.

We developed an analytical approximation for the quantum state with the largest value of $\Gamma$ (\textit{i.e.}, the most superradiant) state of each centriole, modeled as a weighted superposition of the most superradiant states of a set of $104$ $\text{nm}$ (13-spiral) microtubule segments, using the expression
\begin{equation}\label{eq:approxcentriole}
|\phi_0^\text{cent}\rangle \propto \sum_n^N \sum_m^M c_{m,n}\, |\phi_{m,n}^\text{seg}\rangle \,,
\end{equation}
where $|\phi_0^\text{cent}\rangle$ is the bi-orthogonally normalized ~\cite{moiseyev2011non} estimate of the most superradiant centriole state, $M=27$ is the number of microtubules comprising a centriole, and $N$ is the number of $104$-$\text{nm}$ microtubule segments comprising the length of the centriole (\textit{e.g.}, $N=2$ for a $208\,\text{nm}$-long centriole). The indices $m,n$ thus designate the locations of each segment state $|\phi_{m,n}^\text{seg}\rangle$, in effect defining the $27N$ microtubule segments making up the centriole. The real-valued coefficients are defined as
\begin{equation}
c_{m,n} = \sin\left( \frac{\pi\,n}{N+1} \right)\, \sin\left( \frac{2\,\pi\,\lceil{m/3}\rceil}{9} \right) \,,
\end{equation}
where $\lceil x\rceil$ denotes the ceiling function (\textit{i.e.}, the nearest integer larger than $x$). Although inexact, this approach allowed us to validate our fit with an estimate of the complex expectation value of the approximate superradiant state at a computational cost of $\mathcal{O}(n^2)$ numerical operations rather than the usual $\mathcal{O}(n^3)$ required to perform the full matrix diagonalization of the Hamiltonian in Eq.~\ref{Eq_Heff_0E}.

To account for the influence of structural disorder, we carried out additional centriole simulations, introducing variability into the values of Trp peak excitation energies to test the ensemble's robustness to static disorder (Fig.~\ref{fig:centriole}, right panel). 
We considered disorder in the on-site energy of each Trp transition dipole using a random uniform distribution of energies within the range given by $E_0\pm W/2$ for each emitter, introducing $W$ as the disorder parameter. The right panel of Fig.~\ref{fig:centriole} shows the scaling of the normalized superradiant enhancement $\max[\Gamma(W)]/\max[\Gamma(0)]$ for $\Gamma(W)=\{\Gamma_j(W)\}$ as a function of on-site disorder $W$. The normalization factor $1/\max[\Gamma(0)]$ is introduced to allow the direct comparison of enhancements of centrioles of varying length $\ell$. As one can see from the bottom panel of Fig.~\ref{fig:centriole}, as the centriole length increases, a larger value of disorder strength is required to degrade the superradiant enhancement by the same amount. This demonstrates that the Trp network can exhibit cooperative robustness to disorder with increasing network size. The disorder parameter used to obtain the inset in the right panel of Fig.~\ref{fig:centriole} was $W=200\,\text{cm}^{-1}$, which corresponds to the Boltzmann energy $\text{k}_\text{B}T$ at approximately $288\,\text{K}$, revealing that order-of-magnitude $\max(\Gamma_j/\gamma)$ enhancements are plausible at physiological temperatures. 

\begin{figure*}
\hspace{-1mm} a) \hspace{53mm} b) \hspace{54mm} c)\\
\includegraphics[width=0.325\linewidth,trim={0 0 0 25},clip]{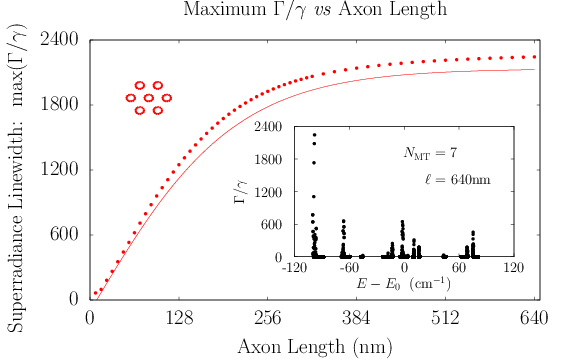}
\includegraphics[width=0.325\linewidth,trim={0 0 0 25},clip]{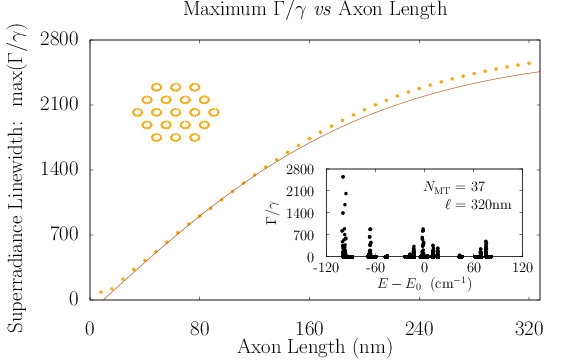}
\includegraphics[width=0.325\linewidth,trim={0 0 0 25},clip]{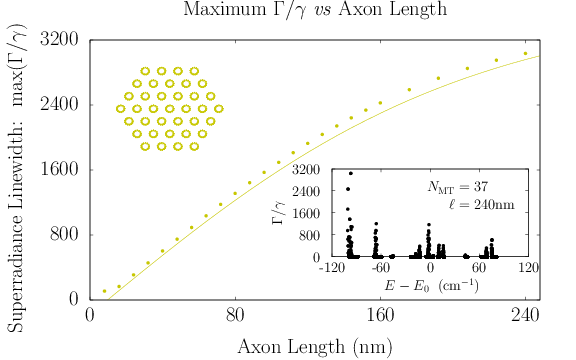} \\
\hspace{-28mm} d) \hspace{53mm} e) \hspace{54mm} f)\\
\includegraphics[width=0.325\linewidth,trim={0 0 0 25},clip]{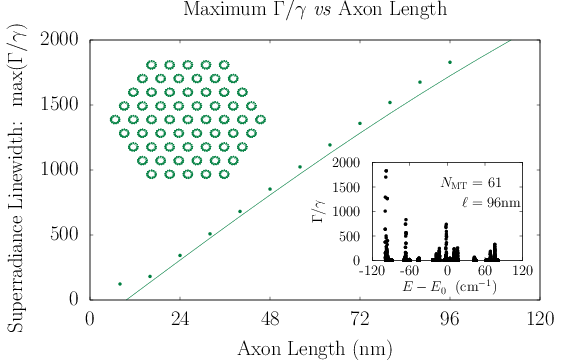}
\includegraphics[width=0.325\linewidth,trim={0 0 0 25},clip]{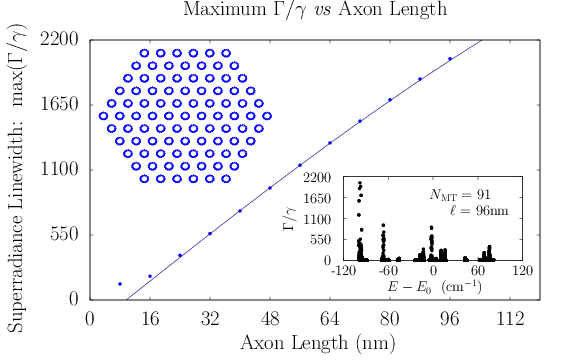}
\includegraphics[width=0.325\linewidth,trim={0 0 0 25},clip]{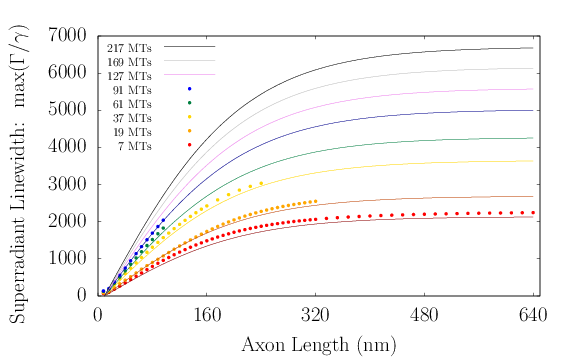}
\caption{\textbf{Neuronal microtubule bundles are predicted to exhibit exceptionally bright states that saturate in superradiant scaling as they approach micron lengths.} Panels a)-e) show linear-linear scale plots of superradiance data $\max(\Gamma\!/\gamma)$ for model axons (MT bundles in hexagonal honeycomb arrangement) of increasing length comprising $N_\text{MT}\in \{7,\! 19,\! 37,\! 61,\! 91\}$ MTs, in that order. In each panel a)-e), the main figure shows the exact $\max(\Gamma_j/\gamma)$ in colored points, approximated by the curve $f^{\text{axon}}_{\text{1JFF}}(\ell, N_\text{MT}) = n_\text{D} (N_\text{MT}/N_0)^{1/d} [(\lambda n_\text{D}/\ell_0)\tanh(\ell/2n_\text{S}\ell_0)-n_\text{S}]$, where $\ell$ is the axon length (in nm) and $N_\text{MT}$ gives the number of MTs it contains. $N_0=7$ is the number of MTs in the smallest hexagonal bundle, $\ell_0=8\,\text{nm}$ is the length along the longitudinal axis of a single MT spiral, $\lambda=280\,\text{nm}$ is the excitation wavelength, $n_\text{D}=8$ is the number of Trp \La~transition dipoles per tubulin dimer shown in Fig. 1a, $n_\text{S}=13$ is the number of tubulin dimers per MT spiral shown in Fig. 1b, and $d=3$ is the spatial dimension. The axon cross-section is shown inset in the upper left of each panel a)-e), and the complex spectrum ($\Gamma/\gamma$ \textit{vs.} $E-E_0$) for the longest exactly-solved axon Trp architecture in the lower right of each panel. Panel f) summarizes plots of the superradiance data $\max(\Gamma_j/\gamma)$ and extrapolates to larger length scales using the analytical function above. The legend of panel f) reflects the color scheme exhibited across panels a)-e). Red, orange, yellow, green, and blue solid curves reflect fits for $N_\text{MT}=$ 7, 19, 37, 61, and 91, respectively, whereas violet, grey, and black curves predict data for axonal MT bundles with $N_\text{MT}=$ 127, 169, and 217, respectively.} 
\label{fig:bigaxon} 
\end{figure*}

\section{Simulations of microtubule bundles in neuronal axons}
\textit{Axons} in neurons can extend vast distances limited only by the scale of the organism \cite{baas2016stability}, ranging from millimeters to meters or longer in mammals, and varying from hundreds of nanometers to microns in diameter \cite{prokop2020cytoskeletal}. Inside an axon, wall-to-wall MT spacings most frequently range from $20\,\text{nm}$ to $30\,\text{nm}$~\cite{chen1992projection}. We simulated Trp networks in  
hexagonal bundles of MTs, spaced 
$50\,\text{nm}$ center-to-center (corresponding to a wall-to-wall MT spacing of $\sim25\,\text{nm}$), with bundle diameters ranging from $\sim 100\,\text{nm}$ to $\sim 0.5\,\mu\text{m}$ and containing 7, 19, 37, 61, or 91 MTs (Fig.~\ref{fig:bigaxon}). 
The resulting superradiance data $\max(\Gamma_j/\gamma)$ data were well-approximated by a set of curves as a function of merely two variables $N_\text{MT}$ and $\ell$, where $N_\text{MT}$ is the number of bundled MTs and $\ell$ is the bundle length, with no free parameters. This allowed us to extrapolate our predictions of $\max(\Gamma_j/\gamma)$ to larger and longer MT bundles (panel~\ref{fig:bigaxon}f), with projected enhancements approaching $\sim7000$. Like that of the centriole, the spectra of these bundles span a range of about $E_0 \pm 100\,\text{cm}^{-1}$, with similar energy band structure in the absence of disorder. For numerical simulations, the Hilbert space dimensions were limited in each case by the available computational resources, which is why the maximum axon lengths diminish as one proceeds from panel~\ref{fig:bigaxon}a to panel~\ref{fig:bigaxon}e. Overall, the results we have obtained from simulations of various hierarchical Trp architectures present the prospect of collective and cooperative UV excitonic states in biological media with different characteristic superradiant maxima and subradiant minima (Table~\ref{TableScales}).

\section{Robustness of quantum yield to disorder} 

In this section we analyze the robustness to static disorder of the quantum yield (QY) enhancements presented in the main text.  Here we consider how time-independent fluctuations of the Trp excitation energies, which are commonly attributed to interaction with distinct local environments, can affect the QY dependence on the Trp network size. 
To account for the influence of structural disorder, we considered time-independent fluctuations in the on-site energies of each Trp transition dipole using a random uniform distribution of energies within the range given by $E_0\pm W/2$ for each emitter, introducing $W$ as the disorder strength parameter. The QY was computed for each realization of disorder and then averaged over ten realizations for each Trp network. Alternatively, one can average over multiple realizations of disorder $W$ the thermally averaged radiative decay rate $\langle \Gamma \rangle_{\text{th}}$ and then use Eq.~2 in the Materials and Methods section of the main text to compute the QY in the presence of structural disorder. We checked that both methods give very similar results. 

\begin{figure*}
\centering
\includegraphics[width=0.5\linewidth,trim={0 0 0 40},clip]
{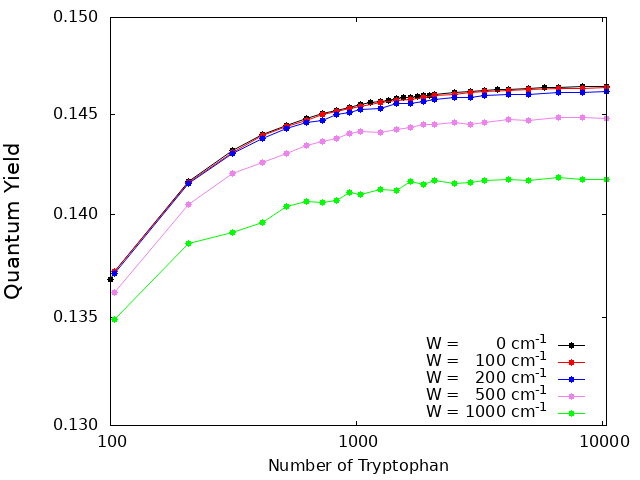}
\includegraphics[width=
0.5\linewidth,trim={0 0 0 30},clip]{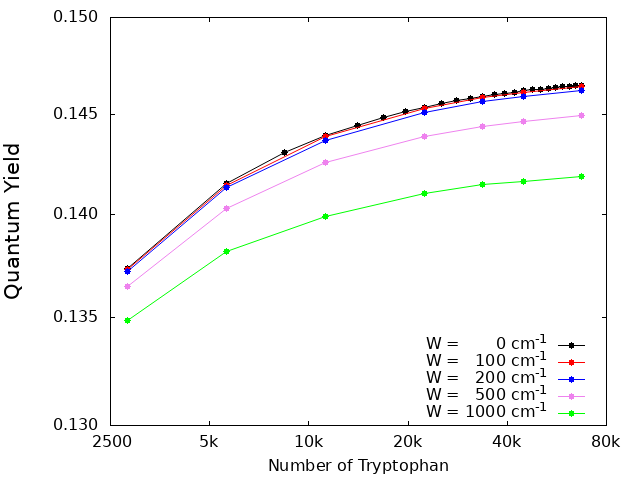}
\caption{\textbf{Quantum yield robustness to static disorder.} The average of the quantum yield 
as a function of microtubule (top panel) and centriole (bottom panel) length, expressed as the number of tryptophans in each architecture, for ten realizations of each static disorder strength $W$ (see legend) is shown. In the top (bottom) panel a single microtuble (centriole) is considered with a maximum length of $\sim800$ ($\sim192$) nm.} 
\label{fig:QYW}
\end{figure*}

In Fig.~\ref{fig:QYW} we show the cases of a single microtubule (top panel) and of a centriole (bottom panel). As one can see, the QY enhancement is extremely robust to static disorder. Indeed, even for a static disorder strength of the same magnitude as room-temperature energy ($200\, \text{cm}^{-1}$), the QY dependence on the microtubule and centriole lengths is basically unaffected. 

The robustness of the QY to structural disorder is remarkable given that such disorder strongly suppresses the superradiant enhancement factor. For instance, in the case of a centriole, the superradiant enhancement goes from $\sim3600$ in the absence of disorder to $\sim20$ for $W=200\,\text{cm}^{-1}$ (see Fig.~\ref{fig:centriole}). Nevertheless the values of the  QY for $W=0\, \text{cm}^{-1}$ and $W=200\, \text{cm}^{-1}$ are very close to each other. The origin of this robustness can be explained as follows: In the presence of static disorder, the superradiant dipole strength gets distributed among  other excitonic states, but states close to the superradiant state in energy will still exhibit most of the dipole strength, if the disorder is not overwhelming. If in the absence of disorder the superradiant state is close to the lowest excitonic state and in the presence of disorder its dipole strength gets distributed within $k_BT$ from it, then the QY is not affected drastically. This means that the QY is a very robust figure of merit for cooperativity. These results strengthen our prediction that an enhancement of the QY will persist even under ambient conditions in biological systems.

\section{Hamiltonian matrix structure for microtubules}

\begin{figure*}[hb]
    a) \hspace{84mm} b) \\
    \includegraphics[width=0.49\textwidth,trim={0 0 0 25},clip]{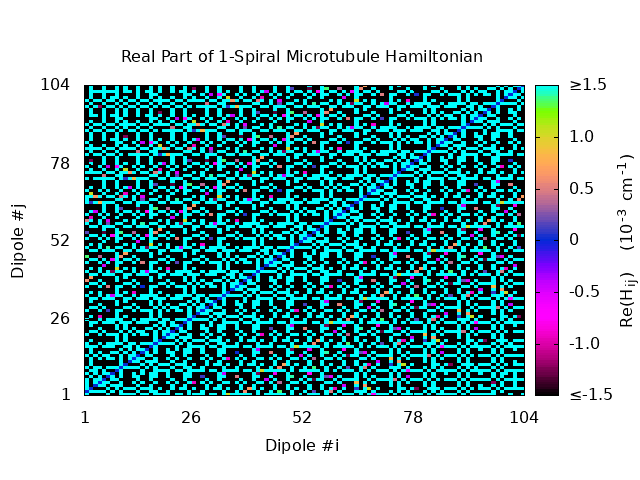}
    \includegraphics[width=0.49\textwidth,trim={0 0 0 25},clip]{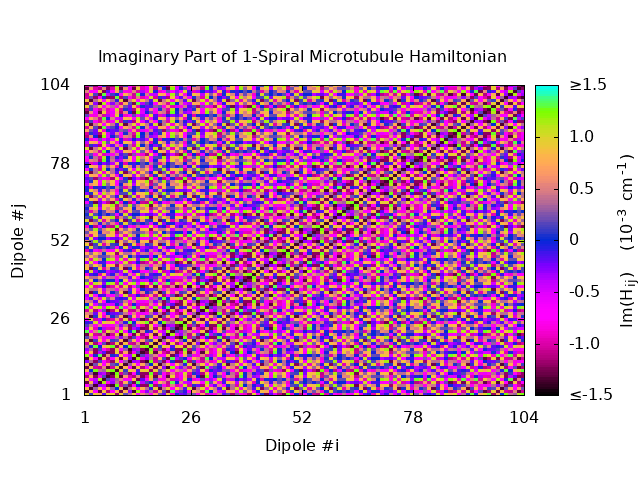} \\
        c) \hspace{84mm} d) \\
     \includegraphics[width=0.49\textwidth,trim={0 0 0 25},clip]{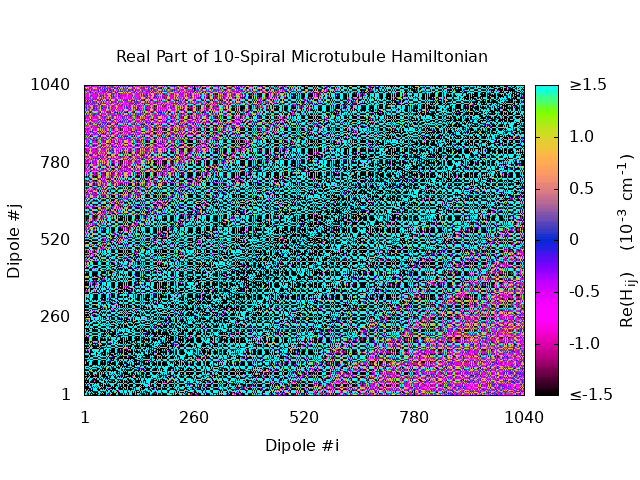}
    \includegraphics[width=0.49\textwidth,trim={0 0 0 25},clip]{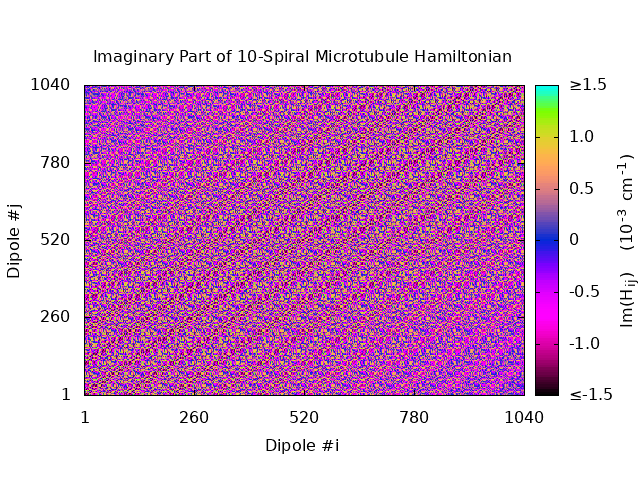}
        e) \hspace{84mm} f) \\
    \includegraphics[width=0.49\textwidth,trim={0 0 0 25},clip]{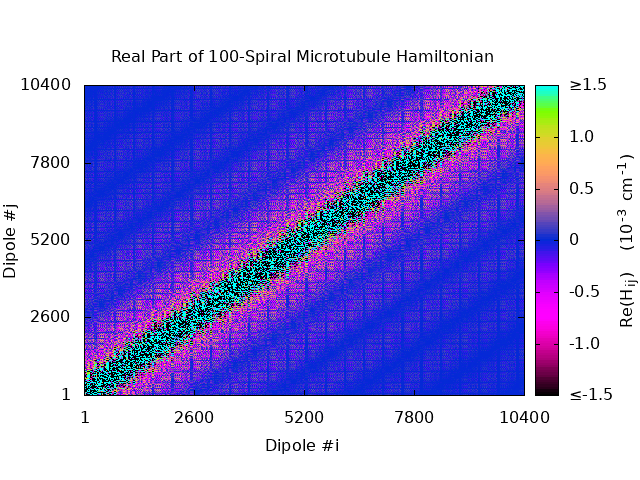}
    \includegraphics[width=0.49\textwidth,trim={0 0 0 25},clip]{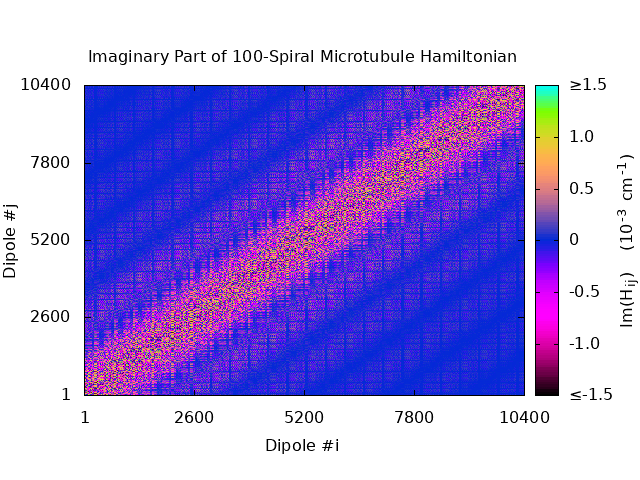}
    \caption{\label{fig:ReImHelements} Panels show real and imaginary matrix elements of Hamiltonians for microtubule (MT) tryptophan networks of varying length: a) $\Re(H_{ij})$ for a 1-spiral MT, b) $\Im(H_{ij})$ for a 1-spiral MT, c) $\Re(H_{ij})$ for a 10-spiral MT, d) $\Im(H_{ij})$ for a 10-spiral MT, e) $\Re(H_{ij})$ for a 100-spiral MT, and f) $\Im(H_{ij})$ for a 100-spiral MT.}
    \label{matrixelements}
\end{figure*}

We have plotted the real and imaginary parts of the inter-chromophore tryptophan couplings from the matrix elements $H_{ij}$ of Eq.~\ref{Eq_Heff_0E}, taking the on-site energies $\Re(H_{jj})$ as zero. Thus, Fig.~\ref{fig:ReImHelements} shows $\Re(H_{ij})$ and $\Im(H_{ij})$ of $\hat{H}$ for a 1-, 10-, and 100-spiral microtubule, respectively.

\end{document}